\documentclass[aps,reprint,groupedaddress,showpacs]{revtex4-1}

\usepackage{euscript,amsmath,amssymb,amsfonts,graphicx,color}
\usepackage{epstopdf}

\newcommand{\e}{{\rm e}}
\renewcommand{\d}{{\rm d}}

\newcommand{\D}{\displaystyle}

\newcommand{\mc}{\mathcal }
\newcommand{\ve}{\varepsilon}
\newcommand{\bph}{{\mathbf \Phi}}
\newcommand{\bphi}{{\boldsymbol \varphi}}
\newcommand{\vp}{\varphi}
\newcommand{\uu}{{\mathbf u}}
\newcommand{\pp}{{\mathbf p}}
\newcommand{\bd}{{\mathbf \Delta}}
\newcommand{\K}{{\mathbf K}}
\newcommand{\W}{{\mathbf W}}
\newcommand{\J}{{\mathbf J}}
\newcommand{\0}{{\mathbf 0}}
\newcommand{\V}{{\mathbf V}}
\newcommand{\vv}{{\mathbf v}}
\newcommand{\Db}{{\mathbf D}}

\begin{document}

\title{Coupling layers regularizes wave propagation in laminar stochastic neural fields}
\author{Zachary P. Kilpatrick}
\email{zpkilpat@math.uh.edu}
\affiliation{Department of Mathematics, University of Houston, Houston, TX 77204}
\date{\today}

\begin{abstract}

We study the effects of coupling between layers of stochastic neural field models with laminar structure. In particular, we focus on how the propagation of waves of neural activity in each layer is affected by the coupling. Synaptic connectivities within and between each layer are determined by integral kernels of an integrodifferential equation describing the temporal evolution of neural activity. Excitatory neural fields, with purely positive connectivities, support traveling fronts in each layer, whose speeds are increased when coupling between layers is considered. Studying the effects of noise, we find coupling also serves to reduce the variance in the position of traveling fronts, as long as the noise sources to each layer are not completely correlated. Neural fields with asymmetric connectivity support traveling pulses whose speeds are decreased by interlaminar coupling. Again, coupling reducers the variance in traveling pulse position, when noise is considered that is not totally correlated between layers. To derive our stochastic results, we employ a small-noise expansion, also assuming inter-laminar connectivity scales similarly. Our asymptotic results agree reasonably with accompanying numerical simulations.

\end{abstract}

\maketitle

\section{Introduction}


There is a growing experimental literature describing the detailed functional architecture of large scale neuronal networks \cite{sporns04}. Much recent development is due to innovative techniques in neural recording such as voltage sensitive dye \cite{ferezou07}, high capacity multielectrodes \cite{nicolelis03}, and optogenetics \cite{zhang10}. Modern electrophysiology reaches well beyond classic neuroanatomical approaches, and details about the complex organization of brain networks in the brain are coming to light \cite{bullmore09}. One particular organizational motif shown to be important for sensory processing is a layered, or laminar, organization of cortical tissue, identified decades ago in visual cortex \cite{hubel77}. An important component of this type of architecture is that synaptic connections between layers have some topographic organization, reflecting recurrent architecture of each local layer \cite{felleman91}. Laminar architecture has now been identified as an important part of motor \cite{weiler08}, somatosensory \cite{shepherd05}, and spatial memory \cite{mcnaughton06} processing. This contributes to previous findings, that connectivity across multiple areas of the brain is important for neural computations like working memory \cite{curtis06}, visual processing \cite{bullier01}, and attention \cite{luck97}. 

The structure of network organization strongly influences the wide variety of spatiotemporal activity patterns observed throughout the brain \cite{wang10}. Strong local recurrent excitation can reenforce stimulus tuning of local assemblies of neurons \cite{song05}, elevating the local response to external inputs \cite{ferster00}, even allowing neural activity to persist seconds after after a stimulus is removed \cite{goldmanrakic95}. In addition, inhibition is known to be dense in many areas of cortex \cite{fino11}, resulting in sharper spatiotemporal responses to sensory stimuli \cite{wehr03}. In addition to relating synaptic polarity to local cortical dynamics, the intricate spatial architecture of many networks in cortex governs those regions' resulting spatiotemporal activity \cite{gilbert96,witter06,vincent07}. For instance, recordings from developing cerebellum reveal that spatial asymmetries in excitatory connectivity can lead to traveling waves of activity \cite{watt09}. Propagating waves of activity have been observed in many sensory cortices \cite{ermentrout01,petersen03,huang04,rubino06}, presumed to amplify or mark the timing of incoming signals. Typically, computational models presume they arise from a combination of recurrent excitation and negative feedback like spike rate adaptation \cite{pinto01} or short term depression \cite{kilpatrick10}.

Therefore, there are a number of established principles concerning how different architectural motifs shape the brain's spatiotemporal activity patterns. We will extend this work by exploring how laminar architecture affects the propagation of activity in stochastic neuronal networks. Recently, we showed that laminar architecture in models of spatial working memory can help stabilize persistent localized activity in the presence of fluctuations \cite{kilpatrick13c}. Persistent activity, in the form of localized bumps, executes a random walk when stochastic fluctuations are considered, but several coupled bumps can cancel much of this noise due to the attractive force between their positions. In this work, similar principles will be demonstrated in stochastic neuronal networks that support traveling waves. Our interest will be in traveling waves that arise from two different mechanisms.

First, we will consider traveling waves in purely excitatory neuronal networks, often used as models for disinhibited cortical slices \cite{richardson05}. Here, the speed of traveling waves is determined by the activity threshold of the network. Second, we will consider traveling waves in asymmetric neuronal networks, previously used as models of direction selectivity \cite{xie02}. The skew of the asymmetry in spatial connectivity determines the speed of traveling waves in these model. Typically, neural field models only consider a single layer of cortical tissue, sometimes separated into distinct excitatory and inhibitory populations \cite{bressloff12}. However, some recent modeling efforts accounted for the multi-laminar structure of cortex, applying them to study interacting bumps \cite{folias11} and binocular rivalry \cite{kilpatrick10b,bressloff12c}. Here, we will combine this approach with a consideration of stochasticity on wave propagation.

There are a number of recent mathematical studies considering how stochasticity affects the formation of spatiotemporal patterns in neuronal networks. Turing patterns \cite{hutt07}, traveling fronts \cite{bressloff12b}, and stationary bumps \cite{kilpatrick13} can all be analyzed in stochastic neural fields with the aid of small-noise expansions originally developed to analyze wave propagation in stochastic partial differential equations \cite{armero98}. Such an approach typically results in a diffusion equation for the position of the spatiotemporal activity, but upon considering a neural field with multiple layers, the effective equations are multivariate Ornstein-Uhlenbeck processes instead \cite{kilpatrick13c}. Thus, the small-noise expansion allows one to examine the effects of connectivity between layers, in addition to noise. Since recordings of cortical activity are becoming substantially more detailed \cite{nicolelis03,zhang10}, the time is ripe for extending theories of spatiotemporal activity patterns in cortex.

The paper will proceed as follows. In section \ref{mod}, we introduce the models we explore, showing how noise and a multi-laminar structure can be introduced into neural field models \cite{hutt07}. One important point is that the correlation structure of spatiotemporal noise can be tuned in the model, and changing this has non-trivial effects on the resulting dynamics.  We proceed, in section \ref{front}, to show how a combination of interlaminar connectivity along with noise affects the propagation of traveling fronts in an excitatory neural field model. As in \cite{kilpatrick13c}, we are able to derive an effective equation for the position of the front, which takes the form of a multivariate OU process. Finally, we derive similar results for traveling pulse propagation in asymmetric neural fields in section \ref{pulses}.

\section{Laminar neural field model} \label{mod}
We will consider two different models for wave propagation in neural fields. They both take the form of a system of coupled stochastic neural field equations
\begin{subequations}  \label{dual}
\begin{align}
\d u_1(x,t) =& \left[ - u_1 + \sum_{k=1}^{2} w_{1k}*f(u_k) \right] \d t + \ve^{1/2} \d W_1 (x,t), \\
\d u_2 (x,t) = & \left[ - u_2 + \sum_{k=1}^2 w_{2k}*f(u_k) \right] \d t + \ve^{1/2} \d W_2 (x,t),
\end{align}
\end{subequations}
where $u_j(x,t)$ is the neural activity of population $j$ at $x \in \Omega$ at time $t$, and the effects of synaptic architecture are describe by the convolution
\begin{align*}
w_{jk}*f(u_k) = \int_{\Omega} w_{jk} (x-y) f(u_k(y)) \d y,
\end{align*}
for $j,k=1,2$, so the case $j=k$ describes recurrent synaptic connections within a layer and $j \neq k$ describes synaptic connections between layers (interlaminar). The function $w_{jk}(x-y)$ describes the strength (amplitude of $w_{jk}$) and net polarity (sign of $w_{jk}$) of synaptic interactions from neurons with stimulus preference $y$ to those with preference $x$. For our analysis of traveling fronts, we consider positive, even weight functions for $w_{jk}$. In particular, we will take the exponential function
\begin{align}
w_{jk} ( x - y ) = \frac{\bar{w}_{jk}}{2} \e^{- |x-y|}, \label{wexp}
\end{align}
so that $\bar{w}_{jk}$ parametrizes the total strength of connections from population $k$ to $j$. Studying excitatory neural fields in section \ref{front}, we extend the spatial domain to $\Omega = (- \infty, \infty)$. In our analysis of coupled traveling pulses in section \ref{pulses}, we presume the modulation of the recurrent synaptic strength is given by the shifted cosine
\begin{align}
w_{jj}(x-y) = \cos (x-y - \phi_j), \ \ \ \ j = 1,2,  \label{acos}
\end{align}
where $\phi_j$ is the amplitude of the shift in the $j$th layer, and the spatial domain is taken to be a periodic ring $\Omega = [- \pi, \pi]$. On the other hand, interlaminar connectivity will generally be given by the pure cosine function
\begin{align}
w_{jk}(x-y) = \bar{w}_{jk} \cos (x-y), \ \ \ k \neq j.   \label{ccos}
\end{align}
Presuming that (\ref{dual}) along with (\ref{acos}) is meant to model directionally selective network, maintaining isotropic coupling between layers reflects a common spatial mapping in the positions within each layer.

Output firing rates are given by taking the gain function $f(u)$ of the synaptic input, which are typically considered to be sigmoidal \cite{wilson73}
\begin{align}
f(u) = \frac{1}{1 + \e^{- \eta (u - \theta)}},  \label{sig}
\end{align}
and we will often take the high gain limit ($\eta \to \infty$), which allows the explicit computation of quantities of interest \cite{amari77}
\begin{align}
f(u) = H(u - \theta) = \left\{ \begin{array}{cl} 1 & : u > \theta, \\ 0 & : u< \theta. \end{array} \right.  \label{H}
\end{align}
Spatiotemporal noises are described by small amplitude ($\ve \ll 1$) stochastic processes $\ve^{1/2} W_j(x,t)$ that are white in time $\langle \d W_j(x,t) \rangle =0$ and correlated in space
\begin{align*}
\langle \d W_j(x,t) \d W_j(y,s) \rangle &= C_j(x-y) \delta (t-s) \d t \d s, \ \ \ j = 1,2, \\
\langle \d W_j(x,t) \d W_k(y,s) \rangle &= C_c(x-y) \delta (t-s) \d t \d s, \ \ \ j \neq k,
\end{align*} 
describing local and shared noise in either layer, $j=1,2$. To demonstrate our theory, we can examine the effects of the simple case of spatially homogeneous noise $C_j(x-y) = \chi_j$ and $C_c(x-y) = \chi_c$. In addition, we can assume a local spatial correlations have a cosine profile $C_j(x) = \chi_j \cos (x/ \sigma)$, along with correlated noise component with a cosine profile so $C_c(x) = \chi_c \cos (x/\sigma)$. Therefore, in the limit $\chi_c \to 0$, there are no interlaminar noise correlations, and in the limit $\chi_c \to \max ( \chi_1, \chi_2)$, noise in each layer is maximally correlated.

\section{Dual layer excitatory network} \label{front}

\subsection{Coupled front propagation}

To begin we examine a network of two coupled excitatory layers, which individually produce propagating fronts. In the presence of sufficient coupling between layers, we will show their front speeds coincide. This analysis should be contrasted with that in \cite{bressloff12c}, which explored fronts coupled with depressing inhibition as a means of modeling binocular rivalry waves.  While our stochastic analysis will only consider the effects of weak coupling, our analysis of the effect of coupling on speed will study the effect of arbitrarily strong coupling. Before analyzing front solutions, it is useful to look at spatially homogeneous solutions of the system (\ref{dual}) in the presence of purely excitatory connections, as they govern the limiting values of traveling fronts. Thus, considering constant solutions $(u_1(x,t),u_2(x,t)) = (\bar{U_1}, \bar{U_2})$, upon plugging them into (\ref{dual}), we have
\begin{align}
\bar{U_1} &= f(\bar{U_1}) \bar{w}_{11} + f(\bar{U_2}) \bar{w}_{12} \nonumber \\
\bar{U_2} &= f(\bar{U_2}) \bar{w}_{22} + f(\bar{U_1}) \bar{w}_{21}, \label{frhomsol}
\end{align}
where $\bar{w}_{jk} = \int_{- \infty}^{\infty} w_{jk}(x) \d x \geq 0$ since $w_{jk}$ are generally positive, even functions. In the limit $\bar{w}_{12}, \bar{w}_{21} \to 0$ and $f$ is a sufficiently steep sigmoid (\ref{sig}) , it can be shown that each equation in (\ref{frhomsol}) will have three roots \cite{ermentrout93}. The largest ($\bar{U}_{1+}$ and $\bar{U}_{2+}$) and smallest ($\bar{U}_{1-}$ and $\bar{U}_{2-}$) of these constitute the boundary conditions of corresponding traveling wave solutions. Thus, as $\bar{w}_{12}, \bar{w}_{21}$ are increased from zero, we expect this fact to still hold over a substantial range of parameters. 

\begin{figure}
\begin{center} \includegraphics[width=8cm]{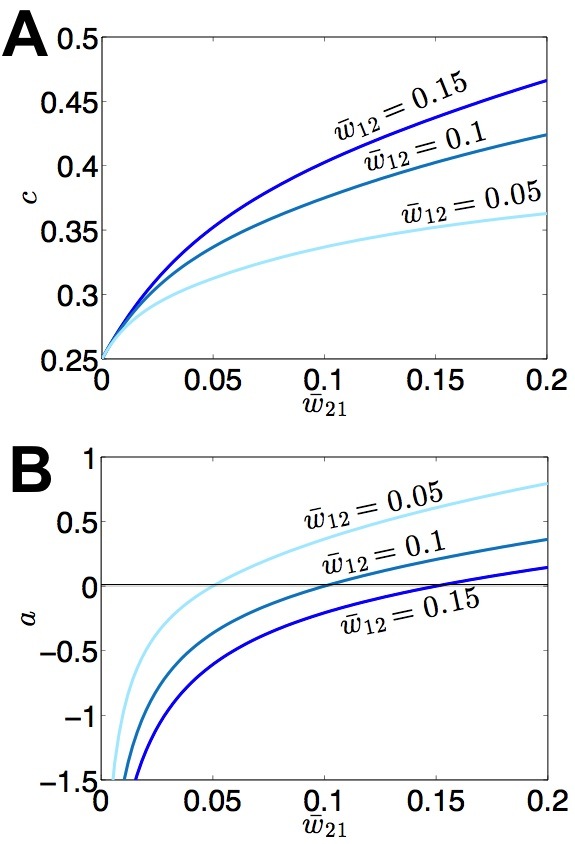} \end{center}
\caption{{\bf A} Speed $c$ and {\bf B} position parameter $a$ of coupled traveling fronts (\ref{frsol1}) and (\ref{frsol2}) as determined by the implicit system (\ref{frspeed1}) and (\ref{frspeed2}). Notice $a=0$ when $\bar{w}_{21}=\bar{w}_{12}$. Other parameters $\bar{w}_{12} = 0.1$ and $\theta=0.4$. }
\label{cupfronts}
\end{figure}

Thus, we seek to construct coupled traveling front solutions to (\ref{dual}) by converting the system to the traveling coordinate frame $\xi = x - ct$, where the wave speed $c$ is yet to be determined. Violations of this assumption will be bifurcations from coupled traveling front solutions. Thus, traveling fronts take the form $(u_1(x,t),u_2(x,t)) = (U_1(\xi),U_2(\xi))$. The translation invariance of the system allows us to set the leading edge of the first front to be at $\xi = 0$ to ease calculations, so they satisfy
\begin{align}
-c U_1'( \xi ) &= - U_1 ( \xi) + w_{11}*f(U_1) + w_{12}*f(U_2), \label{tfront1} \\
- c U_2'(\xi) &= - U_2( \xi ) + w_{22}*f(U_2) + w_{21}*f(U_1),  \label{tfront2} 
\end{align}
where the convolution $*$ is over $\Omega = ( - \infty, \infty)$ with the boundedness conditions $\lim_{\xi \to \pm \infty} U_1 ( \xi ) = \bar{U}_{1 \pm}$ and $\lim_{\xi \to \pm \infty} U_2 ( \xi ) = \bar{U}_{2 \pm}$. The set of equations (\ref{tfront1}) and (\ref{tfront2}) could be solved using shooting methods for an arbitrary choice of nonlinearity $f$ \cite{ermentrout93,pinto01} to specify the wavespeed $c$. However, to demonstrate the relationships between parameters will be proceed by assuming the nonlinearity is Heaviside (\ref{H}). Since we are constructing coupled traveling fronts, there should be a single threshold crossing point for each, yielding the additional conditions $U_1(0) = U_2( a) = \theta$. We can set the threshold crossing point of $U_1$ due to the underlying translation invariance of (\ref{dual}). Note also the threshold crossing point of $U_2$ need not be the same. Thus, we find equations (\ref{tfront1}) and (\ref{tfront2}) become
\begin{align}
-c U_1'( \xi ) &= - U_1( \xi ) + G_1( \xi ) , \label{tfode1} \\
- c U_2'( \xi ) & = - U_2 ( \xi ) + G_2 ( \xi ) , \label{tfode2}
\end{align}
where
\begin{align*}
G_1(x) &= \int_{x}^{\infty} w_{11}(y) \d y + \int_{x-a}^{\infty} w_{12}(y) \d y, \\
G_2 (x) &= \int_{x-a}^{\infty} w_{22} (y) \d y + \int_{x}^{\infty} w_{21} (y) \d y.
\end{align*}
Thus, we can integrate the two equations (\ref{tfode1}) and (\ref{tfode2}) and apply the threshold conditions $U_1(0) = \theta$ and $U_2(a) = \theta$ to yield
\begin{align}
U_1( \xi ) &= \e^{\xi/c} \left( \theta - \frac{1}{c} \int_0^{\xi} \e^{- y/c} G_1( y) \d y \right), \nonumber  \\
U_2 ( \xi ) &= \e^{\xi /c} \left( \theta \e^{-a/c} - \frac{1}{c} \int_a^{\xi} \e^{-y/c} G_2( y) \d y \right).  \label{U12th}
\end{align}
Requiring a bounded solution as $\xi \to \infty $, assuming $c>0$, we have the conditions
\begin{align}
\theta &= \frac{1}{c} \int_0^{\infty} \e^{-y/c} G_1(y) \d y, \nonumber \\
\theta &= \frac{\e^{a/c}}{c} \int_a^{\infty} \e^{-y/c} G_2(y) \d y, \label{thexpr}
\end{align}
so plugging (\ref{thexpr}) into (\ref{U12th}) implies
\begin{align}
U_1( \xi ) &= \frac{1}{c} \int_0^{\infty} \e^{-y/c} G_1(y+ \xi ) \d y, \label{frsol1} \\
U_2( \xi ) &= \frac{1}{c} \int_0^{\infty} \e^{-y/c} G_2(y+ \xi ) \d y. \label{frsol2}
\end{align}
In the case that $w_{jk}$ are all defined as exponential weight distributions (\ref{wexp}) with recurrent weighting $\bar{w}_{11} = \bar{w}_{22} = 1$, the wavespeed $c$ and crossing point $a$ can be related to the threshold $\theta$ and coupling parameters $\bar{w}_{12}$ and $\bar{w}_{21}$ by the implicit system
\begin{align}
\theta &= \frac{1}{2(c+1)} + \bar{w}_{12} {\mc H}(c,-a), \label{frspeed1} \\
\theta &= \frac{1}{2(c+1)} + \bar{w}_{21} {\mc H}(c,+a), \label{frspeed2}
\end{align}
where
\begin{align*}
{\mc H}(c,x) = \left\{ \begin{array}{cl} \frac{\D \e^{-x}}{\D 2 (c+1)} & : x>0, \\  1 + \frac{\D \e^x}{\D 2(c-1)} - \frac{\D c^2 \e^{x/c}}{\D c^2-1} & : x<0. \end{array} \right.
\end{align*}
We solve the system (\ref{frspeed1}) and (\ref{frspeed2}) across a range of value of coupling in Fig. \ref{cupfronts}, showing that the layer receiving more input possesses the leading front. Note, keeping $\bar{w}_{12}>0$ fixed, in the limit $\bar{w}_{21} \to 0$, $a \to - \infty$, so when one layer receives much more excitatory input, its front stays far ahead of the other's. In the case where $\bar{w}_{12} = \bar{w}_{21} = \bar{w}_c$, the system simplifies to a single equation, since the front solution $U_1( \xi ) = U_2( \xi) $ exists, due to reflection symmetry of the full system (\ref{dual}) here. Therefore, $a=0$, so we can write
\begin{align*}
\theta = \frac{1 + \bar{w}_c}{2 (c+1)} \ \ \ \ \Rightarrow \ \ \ \ c = \frac{1 + \bar{w}_c}{2 \theta} - 1,
\end{align*}
so excitatory coupling ($\bar{w}_c>0$) between layers increases the speed $c$ of both fronts. Finally, in the limit $\bar{w}_c \to 0$, there are two decoupled fronts, both with speed $c = 1/(2 \theta) - 1$. This is the limit from which we will build our theory of stochastically driven coupled fronts.

\begin{figure}
\begin{center} \includegraphics[width=8cm]{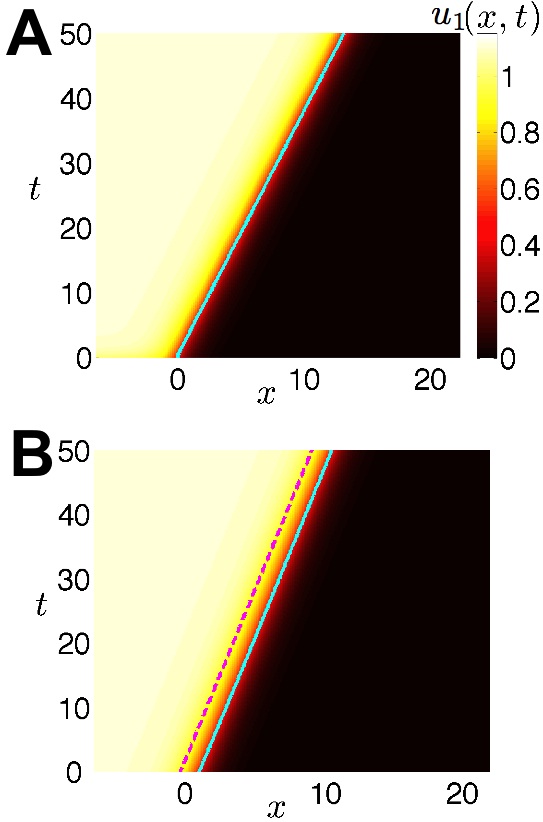} \end{center}
\caption{Evolution of coupled fronts (\ref{frsol1}) and (\ref{frsol2}) in space-time. {\bf A} When $\bar{w}_{12}=\bar{w}_{21} = 0.1$, fronts propagate at the same speed with the same threshold crossing point $x_c(t)$ (solid line), where $u_1(x_c(t),t)=u_2(x_c(t),t) = \theta$. {\bf B} When $\bar{w}_{12}=0.1$ and $\bar{w}_{21}=0.01$, the crossing point $x_1(t)$ of the front in the first layer $u_1(x_1(t),t) = \theta$ (solid) stays ahead of the crossing point $x_2(t)$ (dashed) of the front in the second layer $u_2(x_2(t),t) = \theta$.}
\label{detfrontsim}
\end{figure}

In the limit $\bar{w}_c \to 0$, the fronts (\ref{frsol1}) and (\ref{frsol2}) are neutrally stable to perturbations in both directions. To see this, we consider the perturbed front solutions $u_j(x,t) = U_j(\xi) + \ve U_j'( \xi ) \e^{\lambda t} $, plugging into (\ref{dual}) and truncating to linear order with $w_{11} = w_{22} = w$ and $w_{12} = w_{21} \equiv 0$ to find
\begin{align}
\lambda U_j'(\xi) - c U_j''( \xi ) &= - U_j'( \xi ) + w*[f'(U_j) U_j']  \label{freig}
\end{align}
Differentiating the equations (\ref{tfode1}) and (\ref{tfode2}) and integrating by parts, we find
\begin{align}
c U_j'' - U_j' + w*[f'(U_j)U_j'] = 0,   \label{frdiff}
\end{align}
so the right hand side of (\ref{freig}) vanishes, and $\lambda$ is the only eigenvalue corresponding to translating perturbations. Thus, either front (in layer 1 or 2) is neutrally stable to perturbations that shifts it position in either direction (rightwards or leftwards). We will show in the next subsection, that coupling stabilizes the fronts to perturbations in the opposite directions. Yet, even with coupling, both fronts are neutrally stable to perturbations along the same direction. 

\subsection{Noise-induced motion of coupled fronts}

\begin{figure}
\begin{center} \includegraphics[width=8.5cm]{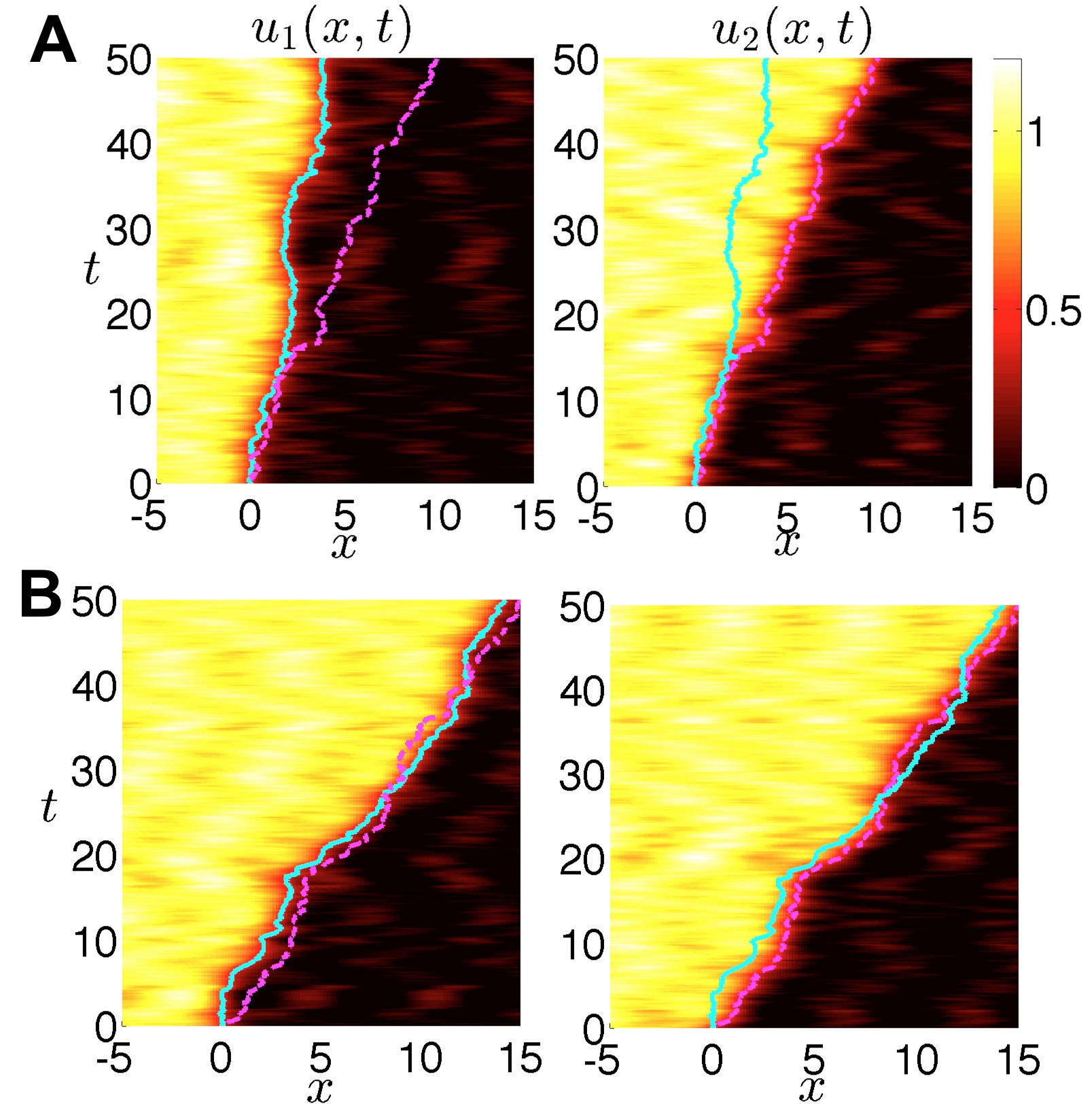} \end{center} \vspace{-4mm}
\caption{{\bf A} Uncoupled fronts $u_1$ and $u_2$ propagating in the dual layer stochastic neural field have leading edges (solid and dashed lines, respectively) that spread apart due to separate sources of noise $\d W_1$ and $\d W_2$. {\bf B} Coupling fronts with connectivity $\bar{w}_{12} = \bar{w}_{21} = 0.05$ keeps noise from spreading fronts very far apart. Coupling is given by exponential weights (\ref{wexp}); other parameters $\theta = 0.4$ and $\ve = 0.01$.}
\label{nosfrontsim}
\end{figure}

Now we consider the effects of small noise on the propagation of fronts in the presence of weak coupling between layers so that $w_{12}, w_{21} = {\mc O}( \ve^{1/2})$ and identical recurrent coupling $w_{11} = w_{22} = w$. To begin, we presume the noise generates two distinct effect in the fronts (see Fig. \ref{nosfrontsim}). First, noise causes both fronts to wander from their paths, while being pulled back into place by the front in the other layer. Each front's displacement from its path will be described by the time-varying stochastic variables $\Delta_1(t)$ and $\Delta_2(t)$. Second, noise causes fluctuations in the shape of both fronts, described by the corrections $\Phi_1(x,t)$ and $\Phi_2(x,t)$. To account for this, we consider the ansatz
\begin{align}
u_1 &= U_1( \xi - \Delta_1(t)) + \ve^{1/2} \Phi_1( \xi - \Delta_1(t), t ) + \cdots  \nonumber \\
u_2 &= U_2( \xi - \Delta_2(t)) + \ve^{1/2} \Phi_2( \xi - \Delta_2(t), t ) + \cdots  \label{franz}
\end{align}
This approach was originally developed to analyze front propagation in stochastic PDE models \cite{armero98}. In stochastic neural fields, it has been modified to analyze wave propagation \cite{bressloff12} and bump wandering \cite{kilpatrick13}. Plugging the ansatz (\ref{franz}) into the system (\ref{dual}) and expanding in powers of $\ve^{1/2}$, we find that at ${\mc O}(1)$, we have the front solution (\ref{frsol1}) and (\ref{frsol2}) when $\bar{w}_{12} = \bar{w}_{21} \equiv 0$. Proceeding to ${\mc O}( \ve^{1/2})$, we find
\begin{align}
\d \bph - {\mc L} \bph = \left( \begin{array}{c} \ve^{-1/2} \d \Delta_1 U_1' + \d W_1 \\ \ve^{-1/2} \d \Delta_2 U_2' + \d W_2 \end{array} \right) + \frac{{\mc K}(x,t)}{\ve^{1/2}},   \label{frplin}
\end{align}
where ${\mc K}(x,t)$ is the $2 \times 1$ vector function
\begin{align*}
{\mc K}= \left( \begin{array}{c} w_{12}*[f(U_2) + f'(U_2) U_2' \cdot ( \Delta_1 - \Delta_2)] \d t \\ w_{21}*[f(U_1) + f'(U_1) U_1' \cdot ( \Delta_2 - \Delta_1) ] \d t \end{array} \right);
\end{align*}
$\bph = ( \Phi_1(\xi , t) , \Phi_2 ( \xi , t))^T$; and ${\mc L}$ is the linear operator
\begin{align*}
{\mc L} \uu = \left( \begin{array}{c} c u'(x) - u(x) + w(x)*[f'(U_1(x))u(x)] \\ c v'(x) - v(x) + w(x)*[f'(U_2(x)) v(x)] \end{array} \right),
\end{align*}
for any vector $\uu = (u(x),v(x))^T$ of integrable functions. Note that the null space of ${\mc L}$ includes the vectors $(U_1',0)^T$ and $(0,U_2')^T$, due to equation (\ref{frdiff}). The last terms in the right hand side vector of equation (\ref{frplin}) arise to due connections between layers. We have linearized them under the assumption $d=\Delta_1 - \Delta_2$ remains small, so $f(U_j(x+d)) \approx f(U_j(x)) -  (-1)^j f'(U_j(x)) U_j'(x) d$, where $j=1,2$. To make sure that a solution to equation (\ref{frplin}) exists, we require the right hand side is orthogonal to all elements of the null space of the adjoint ${\mc L}^*$ which is defined
\begin{align*}
\int_{- \infty}^{\infty} \pp^T {\mc L} \uu \d x = \int_{- \infty}^{\infty} \uu^T {\mc L}* \pp \d x
\end{align*}
for any integrable vector $\pp = (p(x),q(x))^T$. Then,
\begin{align}
{\mc L}^* \pp = \left( \begin{array}{c} -cp'(x) - p(x) + f'(U_1) [w(x)*p(x)] \\ -cq'(x) - q(x) + f'(U_2) [w(x)*q(x)] \end{array} \right). \label{adjfront}
\end{align}
We note that each element of ${\mc L}^*$ is a function of only one element of $\pp$. Therefore, we can decompose the nullspace of ${\mc L}^*$ into two orthogonal elements that take the forms $(\varphi_1,0)^T$ and $(0, \varphi_2)^T$. Thus, we can ensure equation (\ref{frplin}) has a solution by taking the inner product of both sides with the two null vectors to yield
\begin{align*}
\langle \varphi_1, \ve^{-1/2} \d \Delta_1 U_1' + \d W_1 & \\
+ w_{12}*[f(U_2) + f'(U_2)U_2' \cdot ( \Delta_1 - \Delta_2)] \d t \rangle & = 0 \\
\langle \varphi_2, \ve^{-1/2} \d \Delta_2 U_2' + \d W_2 & \\
+ w_{21}*[f(U_1) + f'(U_1) U_1' \cdot ( \Delta_2 - \Delta_1)] \d t \rangle &= 0,
\end{align*}
where we define the inner product $\langle u,v \rangle = \int_{- \infty}^{\infty} u(x) v(x) \d x$. Therefore, the stochastic vector $\bd (t) = ( \Delta_1 (t) , \Delta_2 (t))^T$ obeys the multivariate Ornstein-Uhlenbeck process
\begin{align}
\d \bd (t) = [\J + \K \bd (t)] \d t + \d \W (t)  \label{frred}
\end{align}
where connections between the two layers will slightly alter the mean speed through the term
\begin{align}
\J = \left( \begin{array}{c} \gamma_1 \\ \gamma_2 \end{array} \right) = \left( \begin{array}{c} - \frac{\D \langle \varphi_1 , \ve^{1/2} w_{12}*f(U_2) \rangle}{\D \langle \varphi_1 , U_1' \rangle} \\ -\frac{\D \langle \varphi_2 , \ve^{1/2} w_{21}*f(U_1) \rangle}{\D \langle \varphi_2 , U_2' \rangle}  \end{array} \right)   \label{frJvec}
\end{align}
and pull the positions of both fronts to one another according to the coupling matrix
\begin{align*}
\K = \left( \begin{array}{cc} - \kappa_1 & \kappa_1 \\ \kappa_2 & - \kappa_2 \end{array} \right),
\end{align*}
with
\begin{align*}
\kappa_1 &= \frac{\langle \varphi_1, \ve^{1/2} w_{12}*[f'(U_2)U_2'] \rangle}{\langle \varphi_1, U_1' \rangle}, \\
\kappa_2 &= \frac{\langle \varphi_2, \ve^{1/2} w_{21}*[f'(U_1)U_1'] \rangle}{\langle \varphi_2, U_2' \rangle}.
\end{align*}
Note, in our previous work on stochastic motion of bumps in coupled neural field layers, our effective equation exclusively had deterministic terms of the form in $\K$, due to the solutions $U_j$ being even \cite{kilpatrick13c}. Here, the odd components of the propagating front contribute to the terms in $\J$. Noise is described by the vector $\d \W (t) = ( \d {\mc W}_1, \d {\mc W}_2 )^T$ with
\begin{align*}
\d {\mc W}_1(t) &= - \ve^{1/2} \frac{\langle \varphi_1, \d W_1 \rangle}{\langle \varphi_1 , U_1' \rangle} \\
\d {\mc W}_2 (t) &= - \ve^{1/2} \frac{\langle \varphi_2, \d W_2 \rangle}{\langle \varphi_2, U_2' \rangle}.
\end{align*}
The white noise term $\W$ has zero mean $\langle \W (t) \rangle = \0$ and variance described by pure diffusion so $\langle \W (t) \W^T(t) \rangle = \Db t$ with
\begin{align}
\Db = \left( \begin{array}{cc} D_1 & D_c \\ D_c & D_2 \end{array} \right)   \label{frdcoeff}
\end{align}
where the associated diffusion coefficients of the variance are
\begin{align*}
D_1 &= \ve \frac{\int_{- \infty}^{\infty} \int_{- \infty}^{\infty} \varphi_1(x) \varphi_1(y) C_1(x - y) \d x \d y}{\left[ \int_{- \infty}^{\infty} \varphi_1(x) U_1'(x) \d x \right]^2}, \\
D_2 &= \ve \frac{\int_{- \infty}^{\infty} \int_{- \infty}^{\infty} \varphi_2(x) \varphi_2(y) C_2(x - y) \d x \d y}{\left[ \int_{- \infty}^{\infty} \varphi_1(x) U_1'(x) \d x \right]^2}, 
\end{align*}
and covariance is described by the coefficient
\begin{align*}
D_c = \ve \frac{\int_{- \infty}^{\infty} \int_{- \infty}^{\infty} \varphi_1(x) \varphi_2(y) C_c(x-y) \d x \d y}{\left[ \int_{- \infty}^{\infty} \varphi_1(x) U_1'(x) \d x \right] \left[ \int_{- \infty}^{\infty} \varphi_2(x) U_2'(x) \d x \right]}.
\end{align*}
With the stochastic system (\ref{frred}) in hand, we can show how coupling between layers affects the variability of the positions of fronts subject to noise. To do so, we diagonalize the matrix $\K = \V \Lambda \V^{-1}$ with corresponding matrix
\begin{align*}
\V = \left( \begin{array}{cc} 1 & \kappa_1 \\ 1 & \kappa_2 \end{array} \right),
\end{align*}
which provides us with the decomposition of eigendirections along which the fronts move. The eigenvalue $\Lambda_{11} = \lambda_1 = 0$ corresponds to the neutral stability of the positions $(\Delta_1, \Delta_2)^T$ to translations in the same direction $\vv_1 = (1,1)^T$. The negative eigenvalue $\Lambda_{22} = \lambda_2 = - ( \kappa_1 + \kappa_2)$ corresponds to the linear stability introduced by connections between layers, so the positions $(\Delta_1, \Delta_2)^T$ revert to one another when perturbations translate them in opposite directions $\vv_2 = ( \kappa_1, - \kappa_2)^T$.

With the diagonalization $\K = \V \Lambda \V^{-1}$, assuming $\bd (0) = \0$, the mean $\langle \bd (t) \rangle = \int_0^t \e^{\K(t-s)} \d s \J $, so
\begin{align*}
\langle \bd \rangle = \left( \begin{array}{c} {\mc A}t + {\mc B} \kappa_1 \left( 1- \e^{-(\kappa_1 + \kappa_2)t} \right) \\ {\mc A}t - {\mc B} \kappa_2 \left( 1 - \e^{- ( \kappa_1 + \kappa_2)t} \right) \end{array} \right),
\end{align*}
where ${\mc A} = \frac{\D \gamma_1 \kappa_2 + \gamma_2 \kappa_1}{\D \kappa_1 + \kappa_2}$, ${\mc B} = \frac{\D \gamma_1 - \gamma_2}{\D (\kappa_1 + \kappa_2)^2}$, and we have used the diagonalization $\e^{\K t} = \V \e^{\Lambda t} \V^{-1}$. Since $\lambda_2 = - (\kappa_1 + \kappa_2) < 0$, 
\begin{align*}
\lim_{t \to \infty} \langle \bd (t) \rangle = \left( \begin{array}{c} {\mc A}t + {\mc B} \kappa_1  \\ {\mc A}t - {\mc B} \kappa_2  \end{array} \right),
\end{align*}
so the net mean effect of weak coupling is to slightly increase the wave speed (${\mc A}t$) and potentially alter the relative position of the fronts (${\mc B}$). We would expect this based on the speeding up of fronts observed in our deterministic analysis. Note that if $\gamma_1 = \gamma_2$, then ${\mc B} =0$ and the fronts will have the same mean position. 

To understand the collective effect that noise and coupling has upon relative front positions, we must also study the covariance of the front position vector $\bd (t)$ The formula for the covariance matrix is given by \cite{gardiner04}
\begin{align}
\langle \bd (t) \bd^T (t) \rangle = \int_0^t \e^{\K (t-s)} \Db \e^{\K^T (t-s)} \d s,  \label{frcovint}
\end{align}
where $\Db$ is the covariance coefficient matrix of white noise vector $\W (t)$ given by equation (\ref{frdcoeff}). To compute the integral in (\ref{frcovint}), we use the diagonalization $\K^T = \left( \V^{-1} \right)^T \Lambda \V^T $ so $\e^{\K^T t} = \left( \V^{-1} \right)^T \e^{\Lambda t} \V^T$. By integrating (\ref{frcovint}), we find the elements of the covariance matrix
\begin{align*}
\langle \bd (t) \bd^T (t) \rangle = \left( \begin{array}{cc} \langle \Delta_1(t)^2 \rangle & \langle \Delta_1(t) \Delta_2(t)  \rangle \\ \langle \Delta_1(t) \Delta_2(t) \rangle & \langle \Delta_2(t)^2 \rangle \end{array} \right)
\end{align*}
are
\begin{align}
\langle \Delta_1(t)^2 \rangle &= D_+ t + 2 \kappa_1 r_1(t) + \frac{\kappa_1}{\kappa_2}  r_2(t) \label{frvar1} \\
\langle \Delta_2(t)^2 \rangle  &= D_+ t - 2 \kappa_2 r_1(t) + \frac{\kappa_2}{\kappa_1} r_2 (t)  \label{frvar2} \\
\langle \Delta_1 (t) \Delta_2 (t) \rangle &= D_+ t + ( \kappa_1 - \kappa_2 ) r_1(t) - r_2(t) \label{frcvar} 
\end{align}
where the effective diffusion coefficients are
\begin{align}
D_+ &= \frac{\kappa_2^2 D_1 + 2 \kappa_1 \kappa_2 D_c + \kappa_1^2 D_2}{(\kappa_1 + \kappa_2)^2}  \label{frdiffplus} \\
D_r &= \frac{\kappa_2 D_1 - \kappa_1 D_2 + ( \kappa_1 - \kappa_2) D_c}{(\kappa_1 + \kappa_2)^2} \label{frdiffcross} \\
D_- &= \frac{D_1 - 2 D_c + D_2}{(\kappa_1 + \kappa_2)^2} \label{frdiffminus}
\end{align}
so that $D_+$ and $D_-$ are variances of noises occurring along the eigendirections $\vv_1$ and $\vv_2$. The functions $r_1(t)$, $r_2(t)$ are exponentially saturating 
\begin{align*}
r_1 (t) &= \frac{D_r}{\kappa_1 + \kappa_2} \left[ 1 - \e^{-( \kappa_1 + \kappa_2)t} \right], \\
r_2 (t) &= \frac{\kappa_1 \kappa_2 D_-}{2( \kappa_1 + \kappa_2)} \left[ 1 - \e^{-2 (\kappa_1 + \kappa_2)t} \right].
\end{align*}
We are mainly interested in the variances (\ref{frvar1}) and (\ref{frvar2}) because this will help us to understand how coupling between layers affects the regularity of wave propagation in both layers.

Now, we make a few key observations concerning how coupling affects the position variances (See \cite{kilpatrick13c}, where we analyze the formulae (\ref{frvar1}) and (\ref{frvar2}) in more detail in the context of bump motion in coupled noisy layers, where the main difference was $\J \equiv \0$.). To start, we note that the long term effective diffusion of either front's relative position $\Delta_1(t)$ and $\Delta_2(t)$ will be the same, described by the averaged diffusion coefficient $D_+$, since
\begin{align}
\lim_{t \to \infty} \frac{ \langle \Delta_1(t)^2 \rangle}{t} = \lim_{t \to \infty} \frac{\langle \Delta_2(t)^2 \rangle}{t} = D_+.
\end{align}
The variances $\langle \Delta_j(t)^2 \rangle$ will approach this limit at faster rates as the coupling strengths $\kappa_j$ are increased since other portions of variance decay at a rate determined by $|\lambda_2 | = \kappa_1 + \kappa_2$.

In the case of identical coupling ($w_{12} \equiv w_{21} = w_r$) and noise ($D_1 \equiv D_2 = D_l$), the mean reversion rates will be the same ($\kappa_1 = \kappa_2 = \kappa$) and the terms in (\ref{frdiffcross}) cancel so $D_r = 0$. Thus, the variances will be identical $\langle \Delta_1(t)^2 \rangle = \langle \Delta_2(t)^2 \rangle = \langle \Delta(t)^2 \rangle$ and
\begin{align}
\langle \Delta(t)^2 \rangle = \frac{D_l + D_c}{2} t + \frac{D_l - D_c}{8 \kappa} \left[ 1 - \e^{-4 \kappa t} \right].  \label{varident}
\end{align}
Thus, increases in correlated noise ($D_c$) increase the long-term variance of either front's relative position $\Delta_j$. When noise is entirely shared between layers ($D_l = D_c$) there is no benefit to inter-laminar coupling since $\langle \Delta (t)^2 \rangle = D_lt$ regardless of $\kappa$. If any noise is not shared between layers ($D_c<D_l$), then variance can always be reduced by increasing coupling $\kappa$. Thus, strengthening coupling between two noisy systems can effectively regularize the dynamics. This has been recently shown in the context of coupled noisy oscillators \cite{ly10}. 

\subsection{Calculating stochastic motion of coupled fronts}

We now compute the effective variances (\ref{frvar1}) and (\ref{frvar2}), considering the specific case of Heaviside firing rate functions (\ref{H}) and exponential synaptic weights (\ref{wexp}) with $\bar{w}_{11} = \bar{w}_{22} = 1$. Thus, we can compare our asymptotic results to numerical simulations. First, to compute the front speed corrections $\gamma_1$ and $\gamma_2$, we must calculate the front solutions of the decoupled system \cite{bressloff01,kilpatrick12}
\begin{align}
U_j( \xi) = \left\{ \begin{array}{ll} \theta \e^{- \xi} & : \xi > 0, \\ 1 - \frac{\D (1- 2 \theta)^2}{\D 1- 4 \theta} \e^{\frac{\scriptstyle 2 \theta \xi}{\scriptstyle 1- 2 \theta}} + \frac{\D \theta \e^{\xi}}{\D 1 - 4 \theta} & : \xi < 0, \end{array} \right.  \label{Ujfront}
\end{align}
and their spatial derivatives
\begin{align}
U_j'(\xi) = \left\{ \begin{array}{ll} - \theta \e^{- \xi} & : \xi > 0, \\ - \frac{\D 2 \theta (1- 2 \theta)}{\D 1- 4 \theta} \e^{\frac{\scriptstyle 2 \theta \xi}{\scriptstyle 1- 2 \theta}} + \frac{\D \theta \e^{\xi}}{\D 1 - 4 \theta} & : \xi < 0. \end{array} \right.  \label{Ujpfront}
\end{align}
Now, we can solve explicitly for the null-vectors of ${\mc L}^*$. Plugging (\ref{Ujfront}) and (\ref{Ujpfront}) into (\ref{adjfront}), then we find that each of the two equations in the vector system ${\mc L}^* \bphi = \0$ is
\begin{align}
c \frac{\d \vp_j}{\d \xi} + \vp_j = \frac{\delta( \xi)}{ \theta} \int_{- \infty}^{\infty} w(y) \vp_j (y) \d y, \ \ \ j=1,2,  \label{fradjnul}
\end{align}
where $\bphi = ( \vp_1, \vp_2)^T$. We can integrate (\ref{fradjnul}) to yield
\begin{align}
\vp_j ( \xi ) = - H( \xi ) \e^{- \xi/c}.  \label{franull}
\end{align}
We can then evaluate the integrals in (\ref{frJvec}) to yield
\begin{align}
\gamma_1 = \frac{\ve^{1/2} \bar{w}_{12}}{2 \theta}, \ \ \ \ \ \ \ \  \gamma_2 = \frac{\ve^{1/2} \bar{w}_{21}}{2 \theta},
\end{align}
so as we might expect the fronts will speed up as the strength of inter-laminar connectivity $\bar{w}_{jk}$ is increased. To compute the strength of coupling $\kappa_1$ and $\kappa_2$, we must also compute
\begin{align}
f'(U_j) U_j' = - \delta( \xi ),
\end{align}
in the sense of distributions, so that the coupling terms are given by
\begin{align*}
\kappa_1 = \frac{\ve^{1/2} \bar{w}_{12}}{2 \theta}, \ \ \ \ \ \ \ \ \  \kappa_2 = \frac{\ve^{1/2} \bar{w}_{21}}{2 \theta}.
\end{align*}

We first consider the effect of noise by taking the situation where noise is uncorrelated between layers so $\chi_c = 0$ and $D_c \equiv 0$. Thus, we can simply compute the diffusion coefficients of the local noise in each layer. The simplest choice for spatial correlations to start is globally correlated noise $C_j(x) \equiv  \chi_j$ in each layer $j=1,2$. Then
\begin{align}
D_j = \frac{\ve \chi_j}{\theta^2} \frac{\left[ \int_0^{\infty} \e^{-x/c} \d x \right]^2}{\left[ \int_0^{\infty} \e^{- (1+c) x/c  } \d x \right]^2} = \frac{\ve \chi_j}{4 \theta^4}, \ \ \ j=1,2.  \label{frdjhom}
\end{align}
In addition, we can consider cosine correlations $C_j(x) = \chi_j \cos (x/ \sigma)$ so
\begin{align}
D_j &= \ve \frac{\chi_j \int_{0}^{\infty} \int_{0}^{\infty} \e^{-x/c} \e^{-y/c} \cos((x-y)/\sigma) \d y \d x }{\left[ \theta \int_0^{\infty} \e^{-x/c} \e^{-x} \d x \right]^2} \nonumber \\
& = \frac{\ve \chi_j \sigma^2}{4 \theta^4 (c^2 + \sigma^2)}, \hspace{3cm} j=1,2.  \label{frdjcos}
\end{align}
Note that (\ref{frdjcos}) is an increasing function of $\sigma$ so longer range spatial correlations strength fluctuations' effect on the position of the front via the diffusion coefficients $D_j$. In the limit $\sigma \to 0$, $D_j \to 0$ suggesting that very short range spatial correlations will be insignificant, likely due to averaging by the front's profile. It is also worth considering another nontrivial correlation function $C_j(x) = \chi_j (1+|x|) \e^{-|x|}$, so
\begin{align}
D_j &=  \ve \frac{\chi_j \int_{0}^{\infty} \int_{0}^{\infty} \e^{-x/c} \e^{-y/c}(1+|x-y|)\e^{-|x-y|}\d y \d x }{\left[ \theta \int_0^{\infty} \e^{-x/c} \e^{-x} \d x \right]^2} \nonumber \\
&= \frac{\ve \chi_j (1- \theta)}{\theta^3}, \hspace{3cm} j = 1,2.  \label{frdjexp}
\end{align}
Using any of these effective diffusion coefficients, we can then compute the formulae in (\ref{frvar1}) and (\ref{frvar2}) directly for the case of no noise correlations.

\begin{figure}
\begin{center} \includegraphics[width=8.5cm]{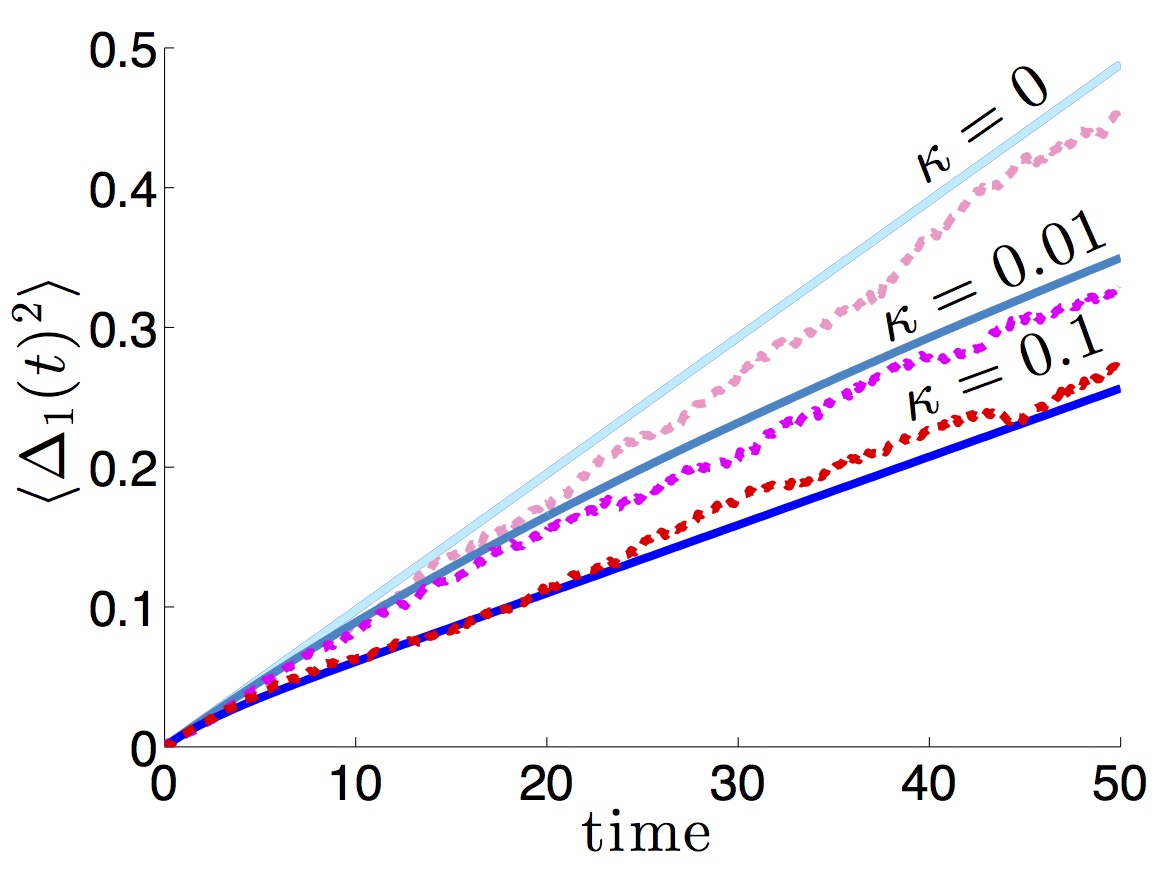} \end{center}
\caption{Effects of spatially homogeneous noise ($C_j(x) = 1$) on propagation of coupled fronts. Theory (solid lines) given by (\ref{frdjhom}) matches numerical simulations (dashed lines) reasonably well. As the strength of identical reciprocal coupling $\kappa_1 = \kappa_2 = \kappa$ is increased, the variance of front position $\langle \Delta_1(t)^2 \rangle$ does not increase as quickly with time. Other parameters are $\theta = 0.4$ and $\ve = 0.001$.}
\label{delfrhfig}
\end{figure}

\begin{figure}
\begin{center} \includegraphics[width=8.5cm]{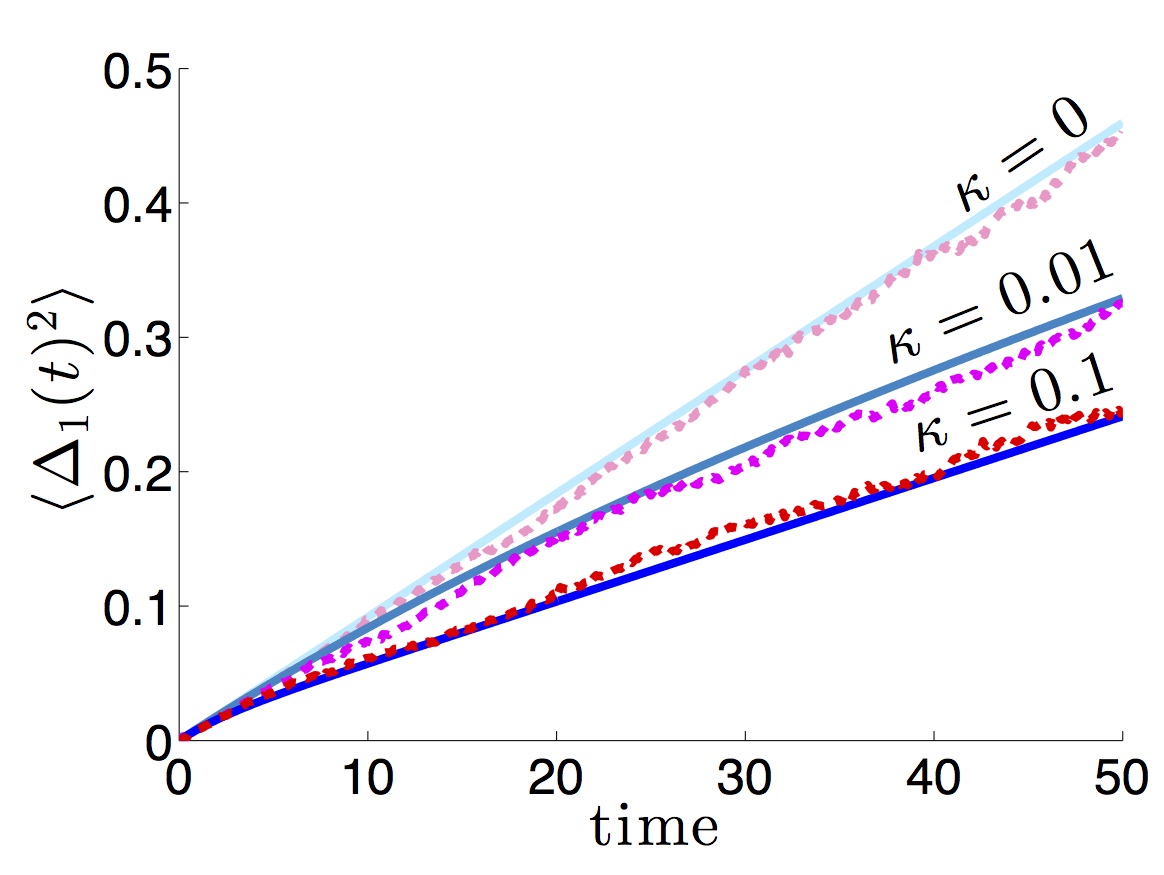} \end{center}
\caption{Effects of cosine correlated noise ($C_j(x) = \cos (x)$) on propagation of coupled fronts. Theory given by (\ref{frdjcos}). As the strength of identical reciprocal coupling $\kappa_1 = \kappa_2 = \kappa$ is increased, the variance of front position $\langle \Delta_1(t)^2 \rangle$ does not increase as quickly with time. Other parameters are $\theta = 0.4$ and $\ve = 0.001$.}
\label{delfrcosfig}
\end{figure}

\begin{figure}
\begin{center} \includegraphics[width=8.5cm]{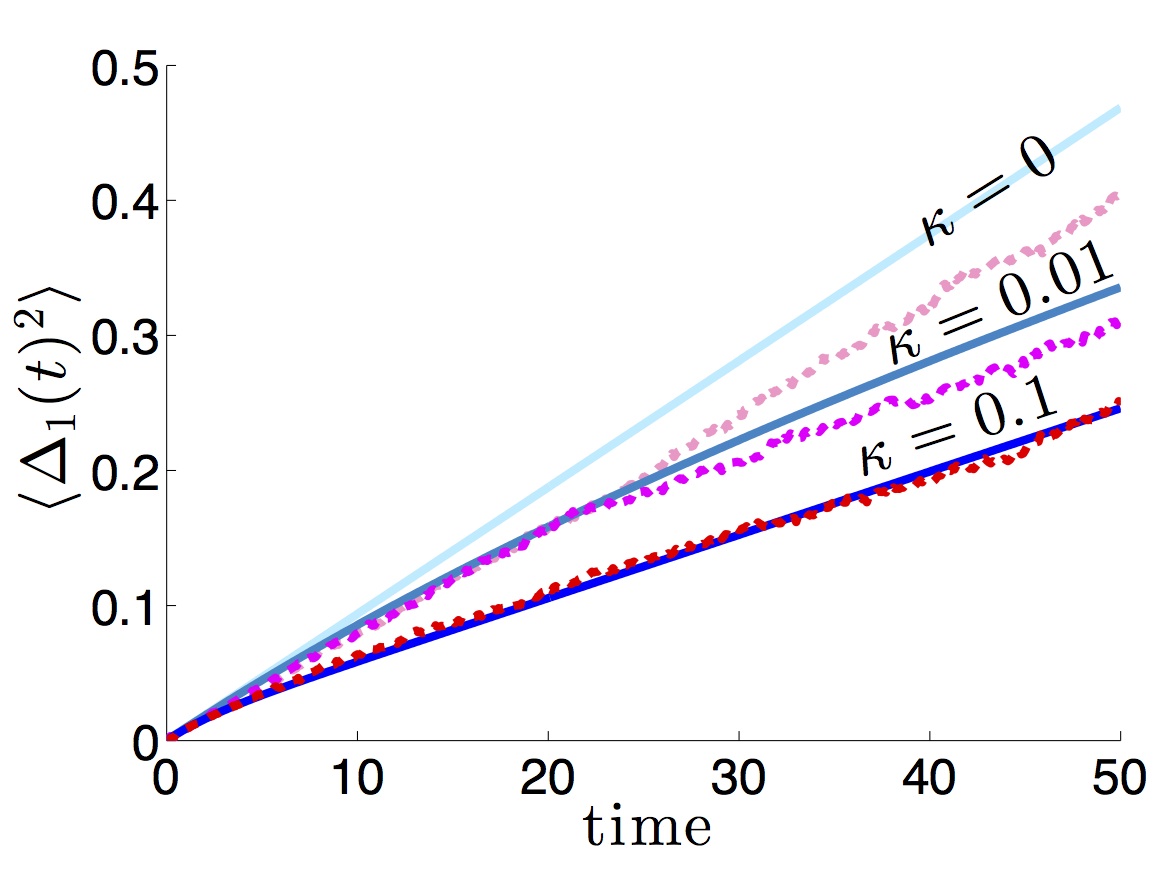} \end{center}
\caption{Effects of exponentially correlated noise ($C_j(x) = (1+|x|)\e^{-|x|}$) on propagation of coupled fronts. Theory given by (\ref{frdjexp}). As the strength of identical reciprocal coupling $\kappa_1 = \kappa_2 = \kappa$ is increased, the variance of front position $\langle \Delta_1(t)^2 \rangle$ does not increase as quickly with time. Other parameters are $\theta = 0.4$ and $\ve = 0.001$.}
\label{delfrexpfig}
\end{figure}

For symmetric connections between areas, $\kappa = \ve^{1/2} \bar{w}_{12}/(2\theta)=\ve^{1/2} \bar{w}_{21}/(2\theta)$, as well as identical noise, $\chi_1 = \chi_2 = 1$, we have $\langle \Delta_1(t)^2 \rangle = \langle \Delta_2(t)^2 = \langle \Delta (t)^2 \rangle$ so that for effective coefficients $D_1 = D_2$, we have 
\begin{align}
\langle \Delta (t)^2 \rangle = \frac{D_j t}{2} + \frac{D_j}{8 \kappa} \left[ 1 - \e^{-4 \kappa t} \right].  \label{delfrhom}
\end{align}
We compare the formula (\ref{delfrhom}) to results we obtain from numerical simulations in Figs. \ref{delfrhfig}, \ref{delfrcosfig}, and \ref{delfrexpfig}.

\begin{figure}
\begin{center} \includegraphics[width=8.5cm]{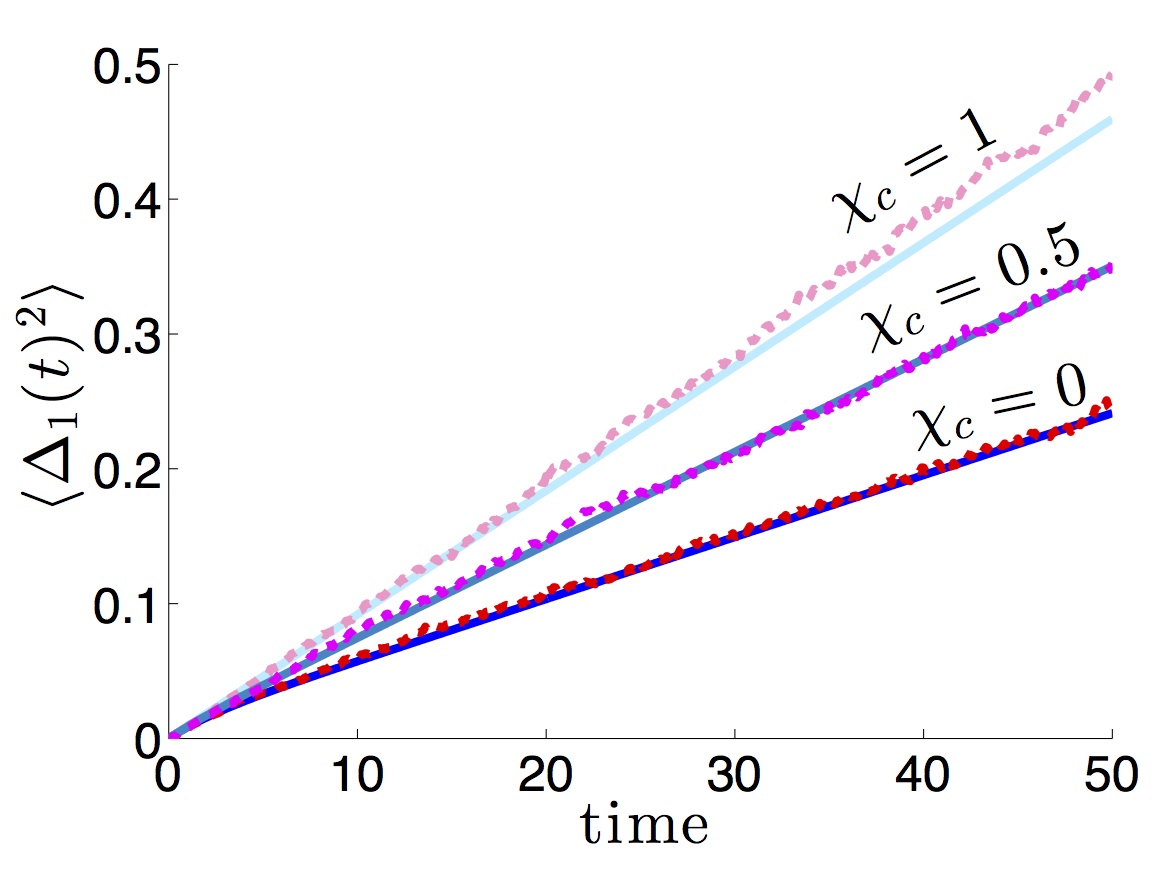} \end{center}
\caption{Effect of correlations between layers, for $C_j(x) = \cos (x)$ and $C_c = \chi_c \cos (x)$, on the propagation of coupled fronts. As the amplitude of noise correlations between layers increases, the effect of reciprocal coupling $\kappa_1 = \kappa_2 = 0.1$ is reduced, as the the variance in front position $\langle \Delta_1(t)^2 \rangle$ scales more quickly all time. In the limit $\chi_c \to 1$, the effects of reciprocal coupling on variance vanish. Other parameters are $\theta = 0.4$ and $\ve = 0.001$.}
\label{delfrcorcos}
\end{figure}

In the case of correlations between layers, so $\chi_c > 0 $, meaning $D_c >0$. In this case, the covariance terms in $D_+$ and $D_-$ are non-zero. We can thus compute the diffusion coefficient associated with correlated noise in the case of cosine correlated noise
\begin{align*}
D_c = \frac{\ve \chi_c \sigma^2}{4 \theta^4 (c^2 + \sigma^2)}.
\end{align*}
In the case of symmetric connections between layers and identical noise, we have $\Delta_1 = \Delta_2 = \Delta$ and for cosine correlated noise
\begin{align}
\langle \Delta (t)^2 \rangle = \frac{(1+ \chi_c) \sigma^2 \ve}{8 \theta^4 (c^2 + \sigma^2)} + \frac{(1-\chi_c) \sigma^2 \ve}{32 \theta^4 (c^2 + \sigma^2) \kappa} \left[ 1 - \e^{-4 \kappa t} \right],  \label{dfrcoscor}
\end{align}
which shows interlaminar connections do not reduce variability as much when noise correlations between layers $\chi_c$ are strong. We demonstrate the accuracy of the theoretical calculation (\ref{dfrcoscor}) in comparison to numerical simulations in Fig. \ref{delfrcorcos}. Essentially, stronger noise correlations between layers diminish the effectiveness of interlaminar connections at reducing front position variance.

\section{Dual ring network}
\label{pulses}

\subsection{Coupled pulse propagation}

We now study another common neural field model framework, asymmetric connectivity that produces traveling pulse solutions \cite{xie02}. To begin, we seek coupled traveling pulse solutions to (\ref{dual}) by constructing solutions in the traveling coordinate frame $\xi = x - ct$ in the absence of noise ($\ve \to 0$), where we will determine the wavespeed $c$ self-consistently. Note, this assumes that the pulses in each layer are locked to one another. We assume this baseline solution and study violations of this assumption as bifurcations from stable coupled traveling pulse solutions. Thus, we assume traveling wave solutions take the form $(u_1(x,t),u_2(x,t)) = (U_1(\xi),U_2(\xi))$. The translation invariance of the system allows us to set the leading edge of the first pulse to be at $\xi = \pi$ to ease calculations. The traveling pulse solutions then satisfy the system
\begin{align}
-c U_1'(\xi) &= -U_1(\xi) + w_{11}*f(U_1) + w_{12}*f(U_2), \label{tpulse1} \\
-c U_2'(\xi) &= -U_2(\xi) + w_{22}*f(U_2) + w_{21}*f(U_1), \label{tpulse2}
\end{align}
where the convolution $*$ is over $\Omega = [- \pi, \pi]$ with the periodic boundary conditions $U_j(- \pi) = U_j( \pi)$ for $j=1,2$. As the system has been projected to a two dimensional set of ordinary differential equations, it can be solved using shooting methods for arbitrary choices of the nonlinearity $f$ \cite{ermentrout93,pinto01} to specify the wavespeed $c$. For purposes of demonstration, we proceed assuming the nonlinearity is a Heaviside (\ref{H}). Since we presume we are constructing coupled traveling pulse solutions, their profiles must cross above and below threshold, yielding the additional conditions $U_1(\pi) = U_1(\pi - a_1) = U_2(b) = U_2(b-a_2) = \theta$. Accounting for the periodicity of the functions $U_1$ and $U_2$ beyond domain $[- \pi, \pi]$, we note that if $b-a_2< - \pi$, the last threshold condition will essentially ensure $U(b-a_2 + 2 \pi ) = \theta$. We can set the front crossing point of $U_1$ to be at $\pi$ due to the underlying translation invariance of the system (\ref{dual}), which we will verify in our linear stability calculations. In addition, note that the leading edge of $U_2$ and width $a_2$ need not be the same as in $U_1$. Therefore, we have the equations (\ref{tpulse1}) and (\ref{tpulse2}) become
\begin{align}
-c U_1' ( \xi) &= - U_1 ( \xi) + G_1( \xi), \label{tpode1} \\
-c U_2'( \xi) &= - U_2( \xi ) + G_2 ( \xi ), \label{tpode2}
\end{align} 
where
\begin{align*}
G_1(x) &= \int_{\pi -a}^{\pi} w_{11} (x - y) \d y + \int_{b-a_2}^b w_{12} (x-y) \d y \\
G_2(x) &= \int_{b-a_2}^b w_{22} (x-y) \d y + \int_{\pi -a}^{\pi} w_{21} (x-y) \d y.
\end{align*}
Thus, we can integrate the two equations (\ref{tpode1}) and (\ref{tpode2}) and apply the threshold conditions $U_1( \pi) = \theta$ and $U_2( b) = \theta$ to yield
\begin{align}
U_1( \xi ) & = \e^{\xi / c} \left( \theta \e^{- \pi/c} - \frac{1}{c} \int_{\pi}^{\xi} G_1(y) \e^{-y/c} \d y \right)  \label{cupuls1} \\
U_2 ( \xi ) &= \e^{\xi / c} \left( \theta \e^{- b/c} - \frac{1}{c}\int_{b}^{\xi} G_2 (y) \e^{-y/c} \d y \right). \label{cupuls2}
\end{align}
By requiring that periodicity holds, $U_j(- \pi ) = U_j( \pi)$ for $j=1,2$, we have
\begin{align*}
2 c \theta \sinh \frac{\pi}{c} &= \int_{-\pi}^{\pi} G_1 (y) \e^{-y/c} \d y \\
2 c \theta \e^{-b/c} \sinh \frac{\pi}{c}  &= \e^{\pi/c} \int_b^{\pi} G_2 (y) \e^{-y/c} \d y \\ &- \e^{-\pi/c} \int_b^{-\pi} G_2 (y) \e^{-y/c} \d y 
\end{align*}
Now, we can generate implicit expressions for the wavespeed $c$, widths $a_1$ and $a_2$, and the position $b$ by applying the remaining threshold conditions $U_1(\pi - a_1) = \theta$ and $U_2(g(a_2)) = \theta$ we have
\begin{align*}
c \theta (\e^{(a_1- \pi)/c} - \e^{- \pi /c})  &=  \int_{\pi-a_1}^{\pi} G_1(y) \e^{-y/c} \d y, \\
c \theta (\e^{(a_2 - b)/c} - \e^{-b/c})  &= \int_{b-a_2}^{b} G_2(y) \e^{-y/c} \d y,
\end{align*}
which can be solved using numerical root finding for a general choice of $w_{jk}$ ($j,k=1,2$).

\begin{figure}
\begin{center} \includegraphics[width=8.5cm]{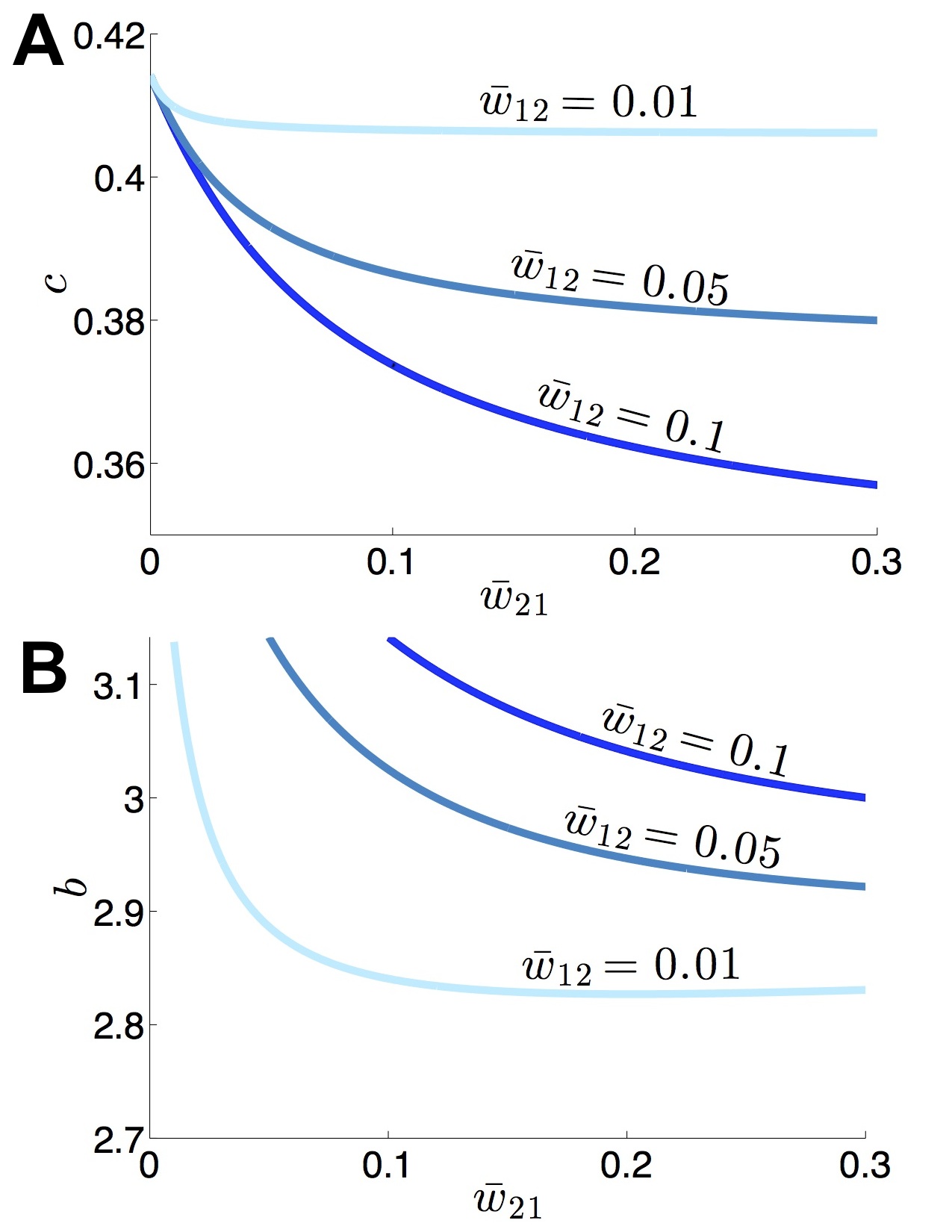} \end{center}
\caption{{\bf A} Speed $c$ and {\bf B} position parameter $b$ of coupled traveling pulses (\ref{cupuls1}) and (\ref{cupuls2}) as determined by the implicit system (\ref{cpth1}) in the a case of asymmetric reciprocal connectivity $\bar{w}_{12} \neq \bar{w}_{21}$, in general. Other parameters $\bar{w}_{12} = 0.1$, $\phi = \pi/8$, and $\theta = 0.4$.}
\label{cpulsasym}
\end{figure}

\begin{figure}
\begin{center} \includegraphics[width=8.5cm]{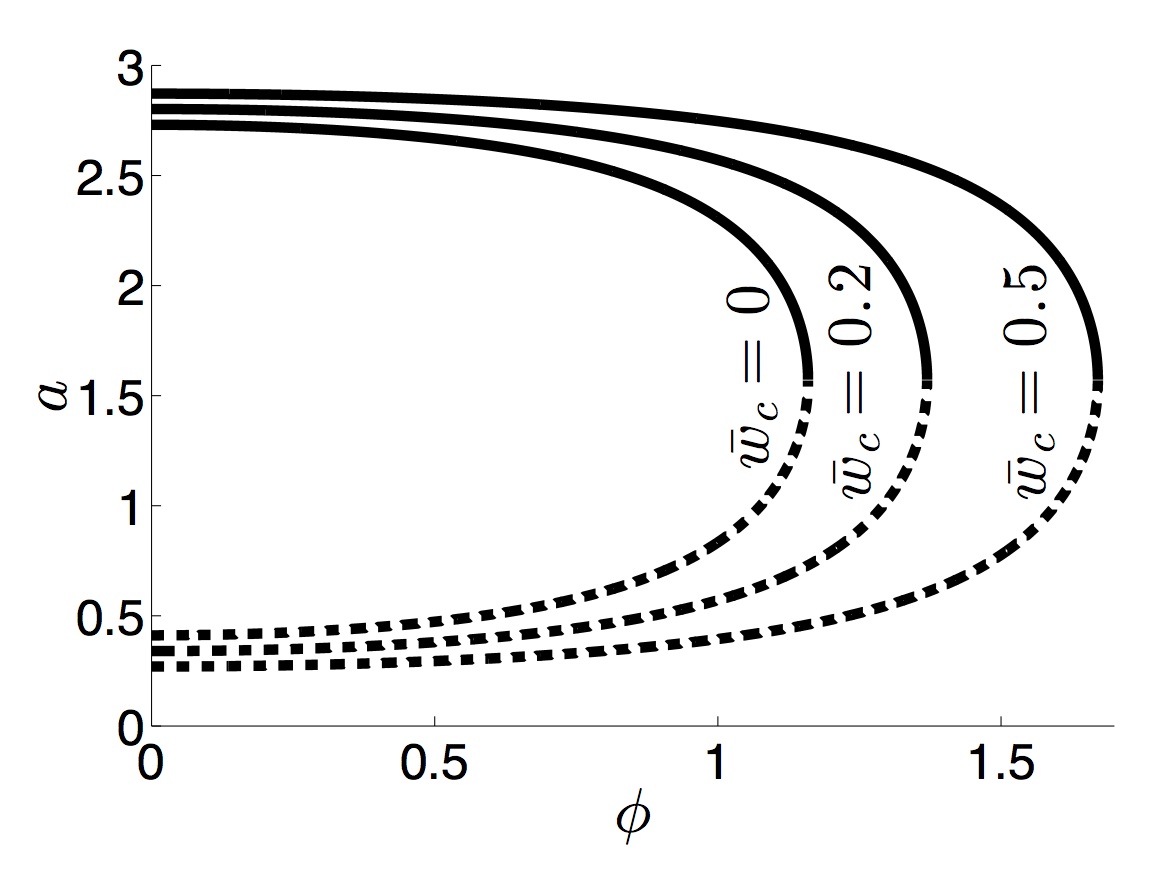} \end{center}
\caption{Pulsewidth $a_1 = a_2 = a$ as a function of the asymmetry $\phi$ of the local weight functions $w_1(x) = w_2(x) = \cos (x - \phi )$ for varying amplitudes of reciprocal symmetric strength $\bar{w}_{12} = \bar{w}_{21} = \bar{w}_c$. Increasing the strength $\bar{w}_c$ shifts the saddle-node bifurcation, at which the stable (solid) and unstable (dashed) branches of pulse solutions, to the right in $\phi$. Other parameter $\theta = 0.4$.}
\label{cpulssym}
\end{figure}

We can compute these expressions in the case where the weight functions are specified by (\ref{acos}) and (\ref{ccos}), so that
\begin{align}
\theta &= \frac{{\mc P}(a_1) + \bar{w}_{12} {\mc Q}(a_2,b)}{c^2+1}, \label{cpth1} \\
\theta &= \frac{ {\mc R}(a_1) + \bar{w}_{12} {\mc S}(a_1,a_2,b) }{c^2+1}, \nonumber \\
\theta &= \frac{{\mc P}(a_2) + \bar{w}_{21} {\mc Q}(a_1,-b) }{c^2+1}, \nonumber \\
\theta &= \frac{{\mc R}(a_2) + \bar{w}_{21} {\mc S}(a_2,a_1,-b)}{c^2+1}, \nonumber
\end{align}
where
\begin{align*}
{\mc P} (x) &= \sin \phi - c \cos \phi + c \cos (x - \phi ) - \sin (\phi - x), \\
{\mc Q} (x,y) &= \sin (y-x) + c \cos y - \sin y - c \cos (x-y), \\
{\mc R} (x) &= \sin (x+ \phi ) - \sin \phi  + c \cos \phi - c \cos (x+ \phi), \\
{\mc S} (x,y,z) &= c \cos (x+z) + \sin (x-y+z) \\
& - \sin (x+z) - c \cos (x-y+z).
\end{align*}
We can solve the system of equations (\ref{cpth1}) numerically to show the effects of varying the coupling $\bar{w}_{12}$ while keeping $\bar{w}_{21}$ fixed. As shown in Fig. \ref{cpulsasym}, increasing the strength $\bar{w}_{12}$ leads to a decrease in wave speed $c$ and a decrease in the position $b$ of the second pulse. In the case of symmetric connectivity $\bar{w}_{12} = \bar{w}_{21} = \bar{w}_c$, the system will simplify to a single equation, specifying the symmetric front solution $U_1( \xi ) = U_2 ( \xi )$. Thus, $b = \pi$ and $a_1 = a_2 = a$, and we can simplify the system to 
\begin{align}
\theta &= \frac{[\cos \phi + c \sin \phi + \bar{w}_c] \sin a}{c^2+1}, \label{cpthsym1} \\
0 & = \frac{(1- \cos a) (\sin \phi - c \cos \phi - c \bar{w}_c)}{c^2 +1}.  \label{cpthsym2}
\end{align}
We can exclude the solution $\cos a = 1$ to (\ref{cpthsym2}), since this will not solve the other equations. Thus, we use the other solution to (\ref{cpthsym2}) to find that
\begin{align*}
c = \frac{\sin \phi}{\cos \phi + \bar{w}_c}
\end{align*}
will be the wave speed. Thus, as opposed to the case of coupled traveling fronts, strengthening connectivity $\bar{w}_c$ here decreases the wave speed. Plugging this into (\ref{cpthsym1}), we find that $\sin a = \theta/\cos \phi + \bar{w}_c$, so 
\begin{align}
a_s &= \pi - \sin^{-1}   \frac{\theta}{\cos \phi + \bar{w}_c}, \label{cpstab} \\ a_u &= \sin^{-1} \frac{\theta}{\cos \phi + \bar{w}_c}  \label{cpunstab}
\end{align}
defines the widths of a coexistent pair of stable (\ref{cpstab}) and unstable (\ref{cpunstab}) coupled traveling pulse solutions. Note that these two branches will coalesce in a saddle-node bifurcation (see \cite{kilpatrick12} for analysis in a single layer network). This bifurcation point is determined by where $\theta = \cos \phi + \bar{w}_c$, as shown in Fig. \ref{cpulssym}.

In the limit $\bar{w}_c \to 0$, the pulses are decoupled, both then having speed $c = \tan \phi$. Pulses will then be neutrally stable to perturbations in both directions. This can be seen by using the same analysis we performed for the excitatory neuronal network that supported fronts. Essentially, perturbations must obey (\ref{freig}), which has an eigenvalue $\lambda = 0$ associated with the eigenfunction $U_j'$ for each layer $j=1,2$. We will now show that coupling layers stabilizes pulses to perturbations that pull them in opposite directions.

\subsection{Noise-induced motion of coupled pulses}

Now, we analyze the effects of weak noise on the propagation of pulses in the presence of reciprocal coupling that is weak ($w_{12},w_{21}={\mc O}(\ve^{1/2})$) and local coupling that is identical ($w_{11} =w_{22} = w$). To start, we presume noise causes each pulse's position to wander, described by stochastic variables $\Delta_1(t)$ and $\Delta_2(t)$, and each pulse's profile fluctuates, described by the stochastic variables $\Phi_1(x,t)$ and $\Phi_2(x,t)$. As in the case of coupled traveling fronts, this is described by the expansion given by the ansatz (\ref{franz}). Plugging this into (\ref{dual}) and expanding in powers of $\ve^{1/2}$, we find the pulse solution at ${\mc O}(1)$ where $\bar{w}_{12} = \bar{w}_{21} \equiv 0$. At ${\mc O}(\ve^{1/2})$, we find the system (\ref{frplin}) with associated linear operator ${\mc L}$, as we found for the excitatory network with fronts. Again, we find that the null space of ${\mc L}$ includes the vectors $(U_1',0)^T$ and $(0,U_2')^T$ due to equation (\ref{freig}). Next, we apply a solvability condition to (\ref{frplin}), where the inhomogenous part must be orthogonal to the nullspace of
\begin{align}
{\mc L}^* \pp = \left(  \begin{array}{c} - c p'(x) - p(x) + f'(U_1) [w(-x)*p(x)] \\ -c q'(x)- q(x) + f'(U_2) [w(-x)*q(x)] \end{array} \right)     \label{puladj}
\end{align}
where $p = (p(x),q(x))^T$. It is important to note that an asymmetric weight function $w(x)$, like (\ref{acos}), leads to a slightly different form for ${\mc L}^*$, now involving terms like $w(-x)*p(x) = \int_{- \pi}^{\pi} w(y-x) p(y) \d y$. Again, we can decomposed the nullspace of ${\mc L}^*$ into two orthogonal elements that take the forms $(\varphi_1,0)^T$ and $(0,\varphi_2)^T$. Rearranging the resulting solvability condition shows that the stochastic vector $\bd (t) = ( \Delta_1(t), \Delta_2(t))^T$ obeys the multivariate Ornstein-Uhlenbeck process
\begin{align}
\d \bd (t) = \left[ \J + \K \bd (t) \right] \d t + \d \W (t)
\end{align}
where connections between the two layers will slightly alter the mean speed through the term
\begin{align}
\J = \left( \begin{array}{c} \gamma_1 \\ \gamma_2 \end{array} \right) = \left( \begin{array}{c} - \frac{\D \langle \varphi_1 , \ve^{1/2} w_{12}*f(U_2) \rangle}{\D \langle \varphi_1 , U_1' \rangle} \\ -\frac{\D \langle \varphi_2 , \ve^{1/2} w_{21}*f(U_1) \rangle}{\D \langle \varphi_2 , U_2' \rangle}  \end{array} \right)   \label{puJvec}
\end{align}
and pull the positions of both fronts to one another according to the coupling matrix
\begin{align*}
\K = \left( \begin{array}{cc} - \kappa_1 & \kappa_1 \\ \kappa_2 & - \kappa_2 \end{array} \right),
\end{align*}
with
\begin{align*}
\kappa_1 &= \frac{\langle \varphi_1, \ve^{1/2} w_{12}*[f'(U_2)U_2'] \rangle}{\langle \varphi_1, U_1' \rangle}, \\
\kappa_2 &= \frac{\langle \varphi_2, \ve^{1/2} w_{21}*[f'(U_1)U_1'] \rangle}{\langle \varphi_2, U_2' \rangle},
\end{align*}
defining the inner product $\langle u,v \rangle = \int_{- \pi}^{\pi} u(x) v(x) \d x$. Noise is described by the vector $\d \W (t) = ( \d {\mc W}_1, \d {\mc W}_2)^T$ with
\begin{align*}
\d {\mc W}_j = - \ve^{1/2} \frac{\langle \varphi_j, \d W_j \rangle}{\langle \varphi_j, U_j' \rangle}, \ \ \ \ j=1,2,
\end{align*}
with mean $\langle \W (t) \rangle = \0$, variance $\langle \W (t) \W^T (t) \rangle = \Db t$, and
\begin{align*}
\Db = \left( \begin{array}{cc} D_1 & D_c \\ D_c & D_2 \end{array} \right)
\end{align*}
with diffusion coefficients
\begin{align*}
D_j = \ve \frac{\int_{- \pi}^{\pi} \int_{- \pi}^{\pi} \varphi_j (x) \varphi_j (y) C_j(x-y) \d x \d y}{\left[ \int_{- \pi}^{\pi} \varphi_j (x) U_j'(x) \d x \right]^2}, \ \ \ j=1,2,
\end{align*}
and covariance described by the coefficient
\begin{align*}
D_c = \ve \frac{\int_{- \pi}^{\pi} \int_{- \pi}^{\pi} \varphi_1 (x) \varphi_2 (y) C_c(x-y) \d x \d y}{\left[ \int_{- \pi}^{\pi} \varphi_1 (x) U_1'(x) \d x \right] \left[ \int_{- \pi}^{\pi} \varphi_2 (x) U_2'(x) \d x \right]}.
\end{align*}
As before, we can diagonalize the system to compute the covariance matrix $\langle \bd (t) \bd (t)^T \rangle$, and we are mainly interested in
\begin{align}
\langle \Delta_1(t)^2 \rangle &= D_+ t + 2 \kappa_1 r_1(t) + \frac{\kappa_1}{\kappa_2} r_2(t) \label{pulvar1} \\
\langle \Delta_2(t)^2 \rangle &= D_+ t - 2 \kappa_2 r_1 (t) + \frac{\kappa_2}{\kappa_1} r_2 (t) \label{pulvar2}
\end{align}
since these give the variance in the positions $\Delta_1$ and $\Delta_2$ of each pulse, which may encode temporal or spatial information. Again, the effective diffusion coefficients are
\begin{align*}
D_+ &= \frac{\kappa_2^2 D_1 + 2 \kappa_1 \kappa_2 D_c + \kappa_1^2 D_2}{(\kappa_1 + \kappa_2)^2} \\
D_r &= \frac{\kappa_2 D_1 - \kappa_1 D_2 + ( \kappa_1 - \kappa_2) D_c}{(\kappa_1 + \kappa_2)^2}  \\
D_- &= \frac{D_1 - 2 D_c + D_2}{(\kappa_1 + \kappa_2)^2}
\end{align*}
and
\begin{align*}
r_1 (t) &= \frac{D_r}{\kappa_1 + \kappa_2} \left[ 1 - \e^{-( \kappa_1 + \kappa_2)t} \right], \\
r_2 (t) &= \frac{\kappa_1 \kappa_2 D_-}{2( \kappa_1 + \kappa_2)} \left[ 1 - \e^{-2 (\kappa_1 + \kappa_2)t} \right].
\end{align*}
As in the excitatory network with fronts, we can note that the long term effective diffusion of both $\Delta_1$ an $\Delta_2$ is $D_+$, and in the case of a symmetric network, variances will be identical and given by (\ref{varident}). Therefore, the main differences will arise in how the particular weight functions (\ref{acos}) and (\ref{ccos}) as well as the shape of the traveling pulses (\ref{cupuls1}) and (\ref{cupuls2}) affects the transfer of noise between layers.

\subsection{Calculating stochastic motion of coupled pulses}

Now, we will compute the variances (\ref{pulvar1}) and (\ref{pulvar2}) considering the specific case of Heaviside firing rate functions (\ref{H}) and cosine synaptic weights (\ref{acos}) and (\ref{ccos}). In particular, we will take $w_{11} = w_{22}$ to compare our asymptotic results to numerical simulations. To compute the pulse speed corrections $\gamma_1$ and $\gamma_2$, we must first calculate the pulse solutions of the decoupled system \cite{xie02,kilpatrick12}
\begin{align}
U_j ( \xi ) = \cos \phi ( \sin \xi - \sin ( \xi + a))  \label{ponsol}
\end{align}
where $a= \pi - \sin^{-1} [ \theta \sec \phi ]$ for the stable pulse. The spatial derivatives
\begin{align}
U_j' ( \xi ) = \cos \phi ( \cos \xi - \cos ( \xi + a)).  \label{ponder}
\end{align}
We can now solve explicitly for the null-vectors of ${\mc L}^*$. Plugging (\ref{ponsol}) and (\ref{ponder}) into (\ref{puladj}) to find the each of the two equations in the vector system ${\mc L}^* \bphi = \0$ is
\begin{align}
c \frac{\d \varphi_j}{\d \xi} + \varphi_j &= C( - \pi ) \delta (\xi + \pi ) + C( \pi - a) \delta ( \xi - \pi + a) \label{padjeqn} \\
C( \xi ) &= \frac{\D  \int_{- \pi}^{\pi} \cos ( y - \xi - \phi ) \varphi_j ( y) \d y}{\D |\cos \phi | [1 - \cos a]}. \nonumber
\end{align}
Using the $2 \pi$ periodicity along with a self-consistency argument, we can solve (\ref{padjeqn}) explicitly to yield \cite{kilpatrick12}
\begin{align*}
\varphi_j ( \xi ) &= \left[ H(\xi + \pi ) + \frac{\coth ( \pi /c ) - 1}{2} \right] \e^{-( \pi + \xi ) / c} \\
& - \left[ H( \xi + a - \pi ) + \frac{\coth ( \pi /c) - 1}{2} \right] \e^{(\pi - a - \xi )/ c}.
\end{align*}
We can then evaluate the integrals in (\ref{puJvec}) to yield
\begin{align*}
\gamma_1 = - \frac{\ve^{1/2} \bar{w}_{12} c}{\cos \phi }, \ \ \ \ \ \gamma_2 = - \frac{\ve^{1/2} \bar{w}_{21} c}{\cos \phi },
\end{align*}
so as predicted by our nonlinear analysis, the pulses will slow down as the strength of inter-laminar connectivity $\bar{w}_{jk}$ is increased. To compute the coupling strengths $\kappa_1$ and $\kappa_2$, we must also compute
\begin{align*}
f'(U_j) U_j' = \delta ( \xi + a - \pi ) - \delta ( \xi - \pi ),
\end{align*}
in the sense of distributions, so that the coupling terms are given
\begin{align*}
\kappa_1 = \frac{\ve^{1/2} \bar{w}_{12}}{\cos \phi }, \ \ \ \ \ \ \ \kappa_2 = \frac{\ve^{1/2} \bar{w}_{21}}{\cos \phi }.
\end{align*}

\begin{figure}
\begin{center} \includegraphics[width=8cm]{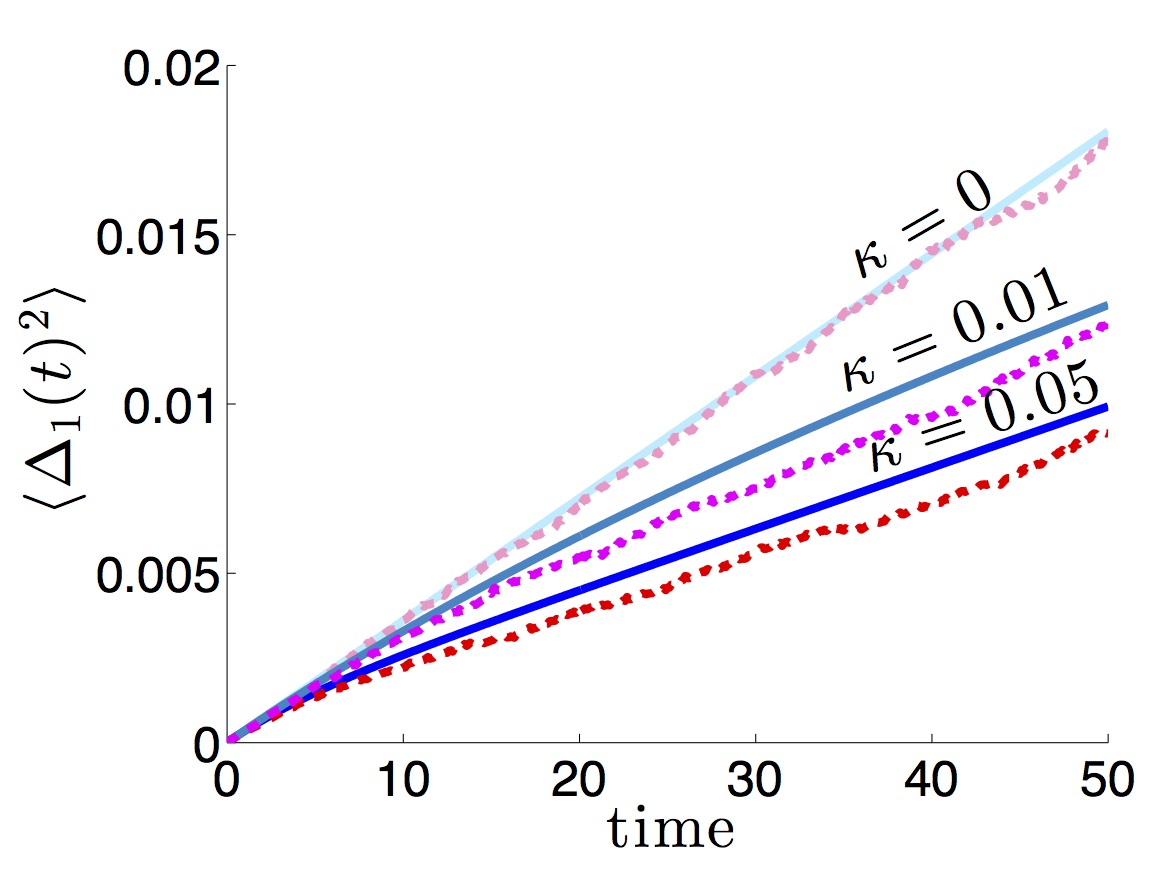} \end{center}
\caption{Effects of cosine correlated noise ($C_j(x) = \cos (x)$) on propagation of coupled pulses. As the strength of identical reciprocal coupling $\kappa_1 = \kappa_2 = \kappa$ is increased, the variance of pulse position $\langle \Delta_1(t)^2 \rangle$ does not increase as quickly with time. Other parameters are $\theta = 0.4$ and $\ve = 0.001$.}
\label{pulvarcos}
\end{figure}

\begin{figure}
\begin{center} \includegraphics[width=8cm]{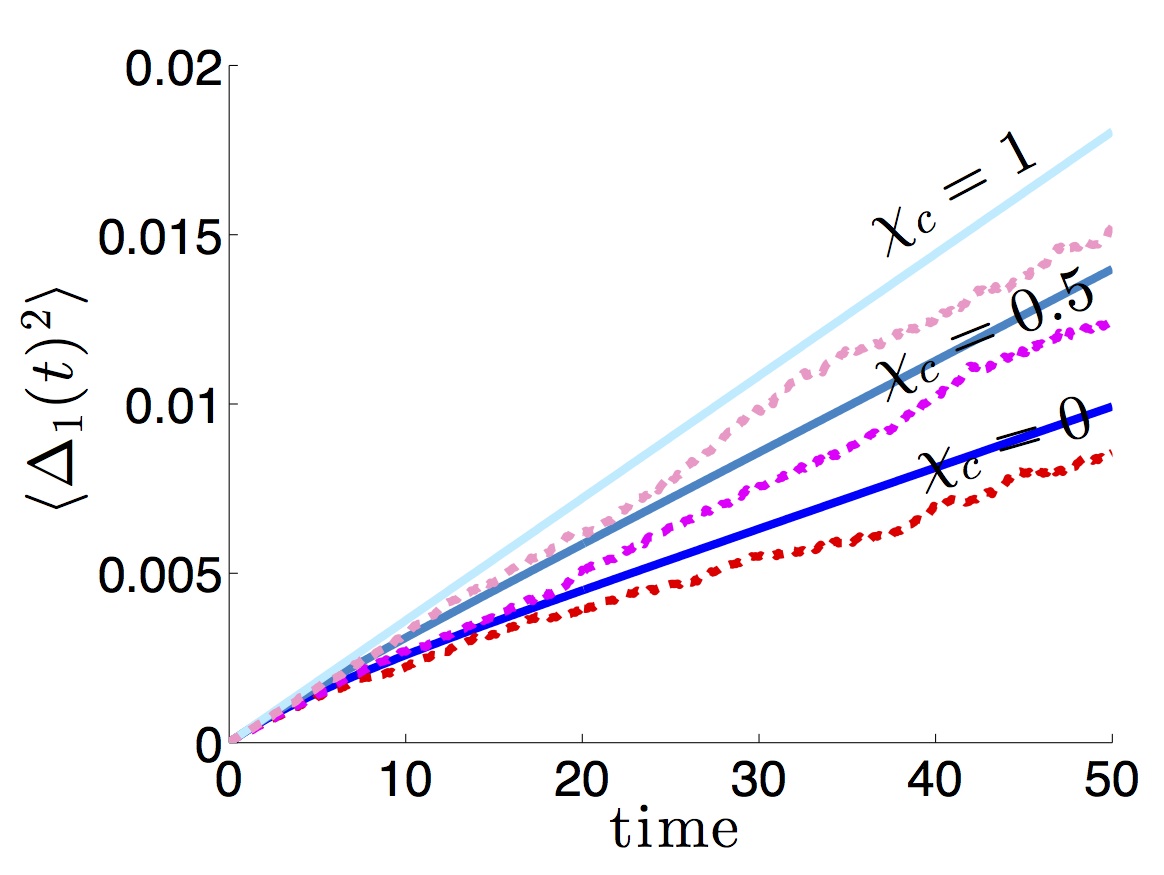} \end{center}
\caption{Effects of cosine correlated noise ($C_j(x) = \cos (x)$) on propagation of coupled pulses when there are noise-correlations between layers ($C_c = \chi_c = \cos (x)$). Increasing the amplitude of correlations $\chi_c$ mitigates the effect of coupling on variance $\langle \Delta_1(t)^2 \rangle$, so that it scales linearly in time in the limit $\chi_c \to 1$. Other parameters are $\theta = 0.4$ and $\ve = 0.001$.}
\label{pulvcorcos}
\end{figure}

Now to consider the effects of noise, we will begin by considering the case where noise is uncorrelated between layers so $\chi_c = 0$ and $D_c \equiv 0$. Thus, we only need to compute the diffusion coefficients in each layer. Starting with the simplest case, globally correlated noise $C_j(x) = \chi_j$ for $j=1,2$, we find
\begin{align*}
D_j = \frac{\ve \chi_j \left[ \int_{- \pi}^{\pi} \varphi_j(x) \d x \right]^2}{4 \cos^4 \phi \sin^2 \phi (1- \cos a)^2} = 0, \ \ j=1,2,
\end{align*}
so globally correlated noise causes no effective perturbation to the positions of the pulses. Thus, we move to considering spatially structure noise correlations given by the cosine $C_j(x) = \chi_j \cos (x)$ so
\begin{align}
D_j &= \ve \chi_j \frac{ \left[ \int_{- \pi}^{\pi} \varphi_j(x) \cos x \d x \right]^2 + \left[ \int_{- \pi}^{\pi} \varphi_j(x) \sin x \d x \right]^2}{4 \cos^4 \phi \sin^2 \phi (1- \cos a)^2} \nonumber \\
&=  \frac{\ve \chi_j }{2 \cos^4 \phi  (1- \cos a)}.  \label{Djcospul}
\end{align}
Using (\ref{Djcospul}) to compute the formulae in (\ref{pulvar1}) and (\ref{pulvar2}), we can compare them with the results obtained for numerical simulations in Fig. \ref{pulvarcos}, specifically using symmetric connectivity $\kappa_1 = \kappa_2 = \kappa$ and noise $\chi_1 = \chi_2 = 1$.

When there are noise correlations between layers ($\chi_c > 0$), covariances in $D_{\pm}$ are non-zero and
\begin{align*}
D_c &=  \frac{\ve \chi_c }{2 \cos^4 \phi  (1- \cos a)}. 
\end{align*}
As was the case for the excitatory network with coupled fronts, by introducing noise correlations between layers, the effects of coupling on variance reduction are lessened. We demonstrate the accuracy of the resulting calculations of symmetric variances $\langle \Delta_1^2 \rangle  = \langle \Delta_2^2 \rangle $ for symmetric connectivity $\kappa_1 = \kappa_2$ and noise $\chi_1 = \chi_2$ in Fig. \ref{pulvcorcos}.

\section{Discussion}

We have demonstrated that reciprocal coupling between layers in multi-layer stochastic neural fields has two main effects on the propagation of neural activity. First, it can alter the mean speed of traveling waves, whether they are fronts or pulses. Second, coupling serves to reduce the variance in wave position in the presence of noise. To demonstrate this, we have derived a multivariate Ornstein-Uhlenbeck process for the position of waves in each layer, under the assumption that the amplitude of noise and connectivity between layers is weak. Variance reduction arises because perturbations that force waves to go in opposite directions decay away due to coupling. Such noise cancelation may arise in various sensory regions in the brain that encode external stimuli with propagating waves and possess multilaminar structure \cite{binzegger04,sakata09}.

There are a number of possible extensions of this work. First of all, we could consider a nonlinear analysis of the laminar stochastic neural field (\ref{dual}) that would account for some of the higher order effects in the variances. This may provide for an even better fit between theory and numerical simulations, especially as the strength of coupling is increased beyond the limit where our linear theory holds. In addition, one common paradigm for generating waves in neural tissue is to provide an external stimulus either in slice \cite{huang04,richardson05} or in vivo \cite{petersen03,rubino06}. One interesting direction would be to examine how external stimulation propagates through a multilaminar network, such as in the networks without space analyzed in \cite{goldman09}. Lastly, there is a great deal of evidence that rats' spatial navigation is encoded by laminar networks in hippocampus and entorhinal cortex \cite{mcnaughton06}. The spatial scales encoded by each layer may vary to generate representations that are nearly redundant, but simply represented at different resolutions \cite{kjelstrup08}. The theory developed here could easily be extended to study laminar networks that represent space at multiple scales and represent two-dimensional space. Our analysis could then lend insight into the neural architecture that leads to the most faithful representation of an animal's present position.

\section*{Acknowledgements} ZPK was funded by NSF-DMS-1311755.

\bibliographystyle{apsrev}
\bibliography{dirsel}

\end{document}